\documentclass[11pt,a4paper]{article}

\usepackage{epsfig}
\usepackage{amsmath}
\usepackage{amssymb}
\usepackage[footnotesize]{caption}
\usepackage{subfigure}
\usepackage{cancel}
\usepackage{cite}
\usepackage{textcomp}
\usepackage{calc}
\usepackage{geometry}
\usepackage{graphicx}
\usepackage{dcolumn}
\usepackage{hyperref}
\usepackage{color}
\usepackage{bbm}

\begin{document}


\noindent September 2013 \hfill \parbox{\widthof{IPMU 13-0188}}{DESY 13-173}

\hfill IPMU 13-0188

\vskip 1.5cm

\begin{center}
{\LARGE\bf A Minimal Supersymmetric Model of\\\smallskip
Particle Physics and the Early Universe}

\vskip 2cm

{\large Wilfried~Buchm\"uller$^a$, Valerie~Domcke$^a$, Kohei~Kamada$^a$, Kai~Schmitz$^b$}\\[3mm]
{\it{
a Deutsches Elektronen-Synchrotron DESY, 22607 Hamburg, Germany\\
b Kavli IPMU (WPI), University of Tokyo, Kashiwa, 277-8583, Japan\\}}
\end{center}

\vskip 1cm


\begin{abstract}

\noindent We consider a minimal supersymmetric extension of the Standard Model,
with right-handed neutrinos and local $B$$-$$L$, the difference between baryon and
lepton number, a symmetry which is spontaneously broken at the
scale of grand unification.
To a large extent, the parameters of the model are determined by 
gauge and Yukawa couplings of quarks and leptons.
We show that this minimal model can successfully account for the earliest
phases of the cosmological evolution:
Inflation is driven by the energy density of a false vacuum of unbroken $B$$-$$L$
symmetry, which ends in tachyonic preheating, i.e.\ the decay of the false vacuum,
followed by a matter dominated phase with heavy $B$$-$$L$ Higgs bosons.
Nonthermal and thermal processes produce an abundance of heavy neutrinos whose
decays generate primordial entropy, baryon asymmetry via leptogenesis and dark
matter consisting of gravitinos or nonthermal WIMPs.
The model predicts relations between neutrino and superparticle masses and a
characteristic spectrum of gravitational waves.
\end{abstract}

\newpage


\tableofcontents


\section{Introduction}


Today, we have the Standard Model of particle physics as well as the $\Lambda$CDM
model of cosmology, which describe a wealth of experimental and
observational data with an accuracy far beyond expectation \cite{Beringer:1900zz}.
On the other hand, despite this success, it is obvious that both standard
models do not represent a final theory.
The symmetry structure of the Standard Model, the smallness of neutrino masses and
the discovery of a Higgs boson~\cite{Aad:2012tfa,Chatrchyan:2012ufa} with a mass in
between the vacuum stability and the triviality bound point towards grand unification
as the next step beyond the Standard Model.
Similarly, the parameters of the $\Lambda$CDM model, the abundance of matter and
dark matter, the apparent cosmological constant and the temperature anisotropies
of the cosmic microwave background (CMB) ask for an explanation which requires physics
beyond the Standard Model.


Supersymmetry is an attractive framework to extrapolate the Standard
Model of particle physics to the energy scale of grand unification,
$\Lambda_{\rm GUT} \sim 10^{16}\,\mathrm{GeV}$.
It also introduces natural dark matter candidates~\cite{Pagels:1981ke,Goldberg:1983nd,Ellis:1983ew}
and scalar fields which can realize inflation, thereby providing an important
link between particle physics and cosmology.
Moreover, neutrino masses require right-handed neutrinos whose
large Majorana masses can account for the tiny masses of the known
neutrinos via the seesaw mechanism.
These Majorana masses break the symmetry $B$$-$$L$, the difference between
baryon and lepton number, and their decays can generate a baryon asymmetry via
leptogenesis~\cite{Fukugita:1986hr}.
Extrapolating the feature of the Standard Model that
all masses are generated by spontaneous symmetry breaking suggests
that also $B$$-$$L$ is a spontaneously broken local symmetry.


Following these arguments, we arrive at a minimal supersymmetric extension
of the Standard Model, which is described by the superpotential
\begin{align}
\label{eq_W}
W = \sqrt{\lambda} \, \Phi \, \left(\frac{v_{B-L}^2}{2} -  \, S_1 S_2 \right)
+ \frac{1}{\sqrt{2}} h_i^n n_i^c n_i^c S_1 + h^{\nu}_{ij} \textbf{5}^*_i n_j^c H_u + W_{\text{MSSM}} \,,
\end{align}
where $S_1$ and $S_2$ are the chiral superfields containing the Higgs
field responsible for breaking $B$$-$$L$, and the $n_i^c$ denote the superfields
containing the charge conjugates of the right-handed neutrinos.
The symmetry-breaking sector of Eq.~\eqref{eq_W}, involving the
superfields $S_1$, $S_2$ and $\Phi$, is precisely the superpotential of
F-term hybrid inflation, with $\Phi$ being a singlet whose scalar component
$\phi$ acts as the inflaton \cite{Copeland:1994vg,Dvali:1994ms}.
$v_{B-L}$ is the scale at which $B$$-$$L$ is broken.
The $B$$-$$L$ charges are $q_S \equiv q_{S_2} = -q_{S_1} =~2$, $q_{\Phi}=0$,
and $q_{n_i^c} = 1$.
$h$ and $\lambda$ denote coupling constants and
$W_{\text{MSSM}}$ represents the MSSM superpotential,
\begin{align}
W_{\text{MSSM}} = h^u_{ij} \textbf{10}_i \textbf{10}_j H_u +
h^d_{ij} \textbf{5}_i^* \textbf{10}_j H_d  \,.
\end{align}
For convenience, all superfields have been arranged in $SU(5)$
multiplets, $\textbf{10}=(q, \, u^c, \, e^c)$ and $\textbf{5}^* =
(d^c, \, l)$, and $i,j = 1,2,3$ are flavor indices.
We assume that the color triplet partners of the electroweak Higgs
doublets $H_u$ and $H_d$ have been projected out.
The vacuum expectation values $v_u = \langle H_u \rangle$ and $v_d = \langle H_d \rangle$
break the electroweak symmetry.
In the following, we will assume large $\tan\beta = v_u/v_d$,
implying $v_d \ll v_u \simeq v_{\text{EW}} = \sqrt{v_u^2 + v_d^2}$. 
We will restrict our analysis to the case of a hierarchical heavy
(s)neutrino mass spectrum, $M_1 \ll M_{2}, M_3$, where $M_i = h_i^n \,v_{B-L}$.
Furthermore, we assume the heavier (s)neutrino masses to be
of the same order of magnitude as the common mass $m_S$ of the
particles in the symmetry-breaking sector, for definiteness we set
$M_{2} = M_3 =  m_S$.
Key parameters of the analysis are then the $B$$-$$L$ breaking scale $v_{B-L}$,
the mass of the lightest of the heavy (s)neutrinos $M_1$, and the effective light
neutrino mass parameter $\widetilde{m}_1$, cf.~\cite{Buchmuller:2012wn},
\begin{align}
\label{eq_v0}
v_{B-L} \simeq \frac{v_{\text{EW}}^2}{\overline{m}_{\nu}} \,, \qquad
M_1 \ll v_{B-L}\,, \qquad \widetilde{m}_1 \equiv \frac{(h^{\nu\dagger}
h^{\nu})_{11} v^2_{\text{EW}}}{M_1} \,.
\end{align}
Here, $\overline{m}_{\nu} = \sqrt{m_2 m_3} \sim 3 \times
10^{-2}\,\mathrm{eV}$ is the geometric mean of the two light neutrino
mass eigenvalues $m_2$ and $m_3$, characterizing the light neutrino mass scale. 
In addition to the chiral superfields, the model also contains a
vector supermultiplet $V$ ensuring invariance under local $B$$-$$L$
transformations and the gravity supermultiplet consisting of the
graviton $G$ and the gravitino $\tilde G$.


In the following sections, we shall show that this Minimal Supersymmetric Model (MSM), 
whose parameters are largely fixed by low-energy experiments,
provides a consistent description of the transition from an inflationary
phase to the hot early universe.
During this `pre- and reheating' process, the matter-antimatter
asymmetry and the dark matter abundance are generated.
Most of our discussion will be based on
Refs.~\cite{Buchmuller:2010yy,Buchmuller:2011mw,Buchmuller:2012wn}.


Our work is closely related to previous studies of thermal
leptogenesis~\cite{Plumacher:1997ru,Buchmuller:2004nz} and nonthermal leptogenesis via 
inflaton decay~\cite{Lazarides:1991wu,Asaka:1999yd,Asaka:1999jb,HahnWoernle:2008pq},
where the inflaton lifetime determines the reheating temperature.
In supersymmetric models with global $B$$-$$L$ symmetry, the scalar 
superpartner $\widetilde{N}_1$ of the lightest heavy Majorana neutrino $N_1$ can play 
the role of the inflaton in chaotic~\cite{Murayama:1992ua,Ellis:2003sq} or 
hybrid~\cite{Antusch:2004hd,Antusch:2010mv} inflation models.
Local $B$$-$$L$ breaking in connection with hybrid, shifted hybrid and
smooth hybrid inflation has been considered in Ref.~\cite{Senoguz:2005bc}.
One of the main motivations for nonthermal leptogenesis has been that the 
`gravitino problem' for heavy unstable
gravitinos~\cite{Weinberg:1982zq,Ellis:1984er,Kawasaki:2004yh,Kawasaki:2004qu,Jedamzik:2006xz}
can be avoided by means of a low reheating temperature.
In the following, we shall assume that the gravitino is either the lightest
superparticle (LSP) or very heavy, $m_{\widetilde{G}} \gtrsim 10~\mathrm{TeV}$.
In the first case, gravitinos, thermally produced at a reheating temperature compatible with 
leptogenesis, can explain the observed dark matter abundance \cite{Bolz:1998ek}. 
For very heavy gravitinos, thermal production and subsequent decay
into a wino or higgsino LSP can yield nonthermal WIMP dark
matter~\cite{Gherghetta:1999sw,Ibe:2004tg,Buchmuller:2012bt}.


The MSM, defined in Eq.~\eqref{eq_W}, postdicts the earliest phases of the cosmological evolution.
The energy density of a false vacuum with unbroken $B$$-$$L$ symmetry drives inflation.
Consistency with the measured amplitude of the temperature anisotropies of the cosmic
microwave background fixes $v_{B-L}$, the scale of $B$$-$$L$
symmetry breaking, to be the GUT scale.
Inflation ends by tachyonic preheating \cite{Felder:2000hj}, i.e.\ the decay of
the false vacuum, which sets the stage for a phase dominated by nonrelativistic
matter in the form of heavy $B$$-$$L$ Higgs bosons.
The further development is described by Boltzmann equations.
Nonthermal and thermal processes produce an abundance of
heavy neutrinos whose decays generate primordial entropy, 
baryon asymmetry via leptogenesis and gravitino dark matter from scatterings
in the thermal bath.
This whole pre- and reheating process is
imprinted on the spectrum of primordial gravitational waves \cite{Buchmuller:2013lra}.
It is remarkable that the initial conditions of the radiation dominated phase are not
free parameters of a cosmological model.
Instead, they are determined by the parameters of a Lagrangian,
which in principle can be measured by particle physics experiments
and astrophysical observations. 
The consistency of hybrid inflation, leptogenesis and dark matter entails
interesting relations between the lightest neutrino mass $m_1$, the gravitino
mass and possibly wino or higgsino masses. 


The paper is organized as follows.
In Sec.~2, we discuss F-term hybrid inflation.
Corrections from supersymmetry breaking lead to a
two-field model which can account for all results deduced from the recently
released PLANCK data.
Sec.~3 deals with tachyonic preheating and
the important topic of cosmic string formation, with emphasis on the
current theoretical uncertainties.
The description of the reheating process by means of Boltzmann equations
and the resulting relations between neutrino masses and superparticle masses
are the subject of Sec.~4.
The predictions of the gravitational wave spectrum due to
inflation, cosmic strings, pre- and reheating are reviewed in
Sec.~5.
Finally, observational prospects are addressed in Sec.~6. 


\section{Inflation}


The superpotential of the MSM, cf.\ Eq.~\eqref{eq_W}, allows for
a phase of F-term hybrid inflation.
For $|\phi| \gtrsim v_{B-L}$, the $B$$-$$L$ Higgs fields are fixed at
zero, $B$$-$$L$ is unbroken and the energy density of the universe
is dominated by the false vacuum energy density, $\rho_0 \simeq (\lambda/4) \, v_{B-L}^4 \equiv V_0$,
generated by the non-vanishing vacuum expectation value (vev) of the auxiliary field
$F_\phi$ and inducing spontaneous supersymmetry breaking.
Here, we briefly review the dynamics and predictions of this inflation model,
with particular focus on the status of F-term hybrid inflation in the light
of the recent PLANCK results~\cite{Planck:2013jfk}.


\subsection{Scalar potential}


At the high energy scales involved in inflation, supergravity corrections to
the Lagrangian become important, resulting in a tree-level scalar F- and D-term
potential given by
\begin{align}
 V^F_{\text{SUGRA}} = e^{K/M_{\rm Pl}^2} \left[ \sum_{\alpha \bar \beta}
 K^{\alpha \bar \beta} \, {\cal D}_\alpha W \,  {\cal D}_{\bar \beta}
 W^* - 3 \, \frac{|W|^2}{M_{\rm Pl}^2} \right] \,, \qquad 
V^D_{\text{SUGRA}} = \frac{1}{2} \,  g^2 \left(\sum_\alpha q_\alpha \,  K_\alpha \, z_\alpha \right)^2 \,, 
\label{eq_Vsugra}
\end{align}
where ${\cal D}_\alpha W = W_\alpha + K_\alpha W/ M_{\rm Pl}^2$; the subscript $\alpha  \, (\bar \alpha)$
denotes the derivative with respect to the (complex conjugate of the) scalar component $z_\alpha$
of the superfield $\Phi_\alpha$ carrying $U(1)$ gauge charge $q_\alpha$.
Moreover, $K^{\alpha \bar \beta}$
is the inverse K\"ahler metric and $M_{\rm Pl} = 2.4 \times 10^{18}$~GeV denotes the reduced Planck mass.
For a canonical K\"ahler potential,
\begin{align}
 K = \sum_\alpha |z_\alpha|^2 \,,
\end{align}
the D-term scalar potential reduces to the expression familiar from global supersymmetry, but
an important supergravity contribution arises from the F-term potential,
$ V^F_\text{SUGRA} \supset |z_\alpha|^2 \rho_0 / M_{\rm Pl}^2 $.
This yields large contributions to the masses of the scalar fields
$z_\alpha$ of the theory.
For the superpotentialm in Eq.~\eqref{eq_W}, this stabilizes the singlet
sneutrinos and the MSSM scalars at a vanishing field value. 
The $B$$-$$L$ Higgs boson masses also obtain various supergravity
contributions.
However, these are suppressed by factors of $(v_{B-L}/M_{\rm Pl})^2$ or $(\phi/M_{\rm Pl})^2$
compared to the leading order terms, which match the result found in global supersymmetry,
\begin{align}
\left(m^S_\pm\right)^2 = \lambda\left(\left|\phi\right|^2 \pm \frac{1}{2}v_{B-L}^2\right)
\,, \quad \left(m_f^S\right)^2 = \lambda \left|\phi\right|^2 \,.
\label{eq_S1S2masses}
\end{align}


The F-term supergravity contribution discussed above does not give a mass term
to the inflaton $\phi$ because, after expanding $e^{K/M_{\rm Pl}^2}$ in Eq.~\eqref{eq_Vsugra},
the term in question is canceled by the corresponding term in $D_\phi W D_{\bar \phi} W^*$.
The leading order supergravity contribution to the inflaton mass thus stems from
the term proportional to $ |\phi|^4 v_{B-L}^4 /M_{\rm Pl}^4$ in the
scalar potential~\cite{Linde:1997sj}.
In addition, since supersymmetry is broken during inflation, we need to take
the one-loop Coleman-Weinberg (CW) potential for the inflaton field into account,
obtained by integrating out the heavy $B$$-$$L$ Higgs bosons,
\begin{align}
 V_{1\ell} = \frac{1}{\rule{0pt}{10pt} 64 \pi^2} \, \text{STr} \left[ M^4
 \left( \ln \left(\frac{M^2}{\rule{0pt}{10pt} Q^2} \right) - \frac{1}{2} \right) \right] 
\simeq \frac{\lambda^2 v_{B-L}^4}{\rule{0pt}{10pt} 64 \pi^2} \left[ \ln
\left( \frac{ 2 |\phi|^2}{\rule{0pt}{10pt} v_{B-L}^2}\right) + {\cal O}
\left( \frac{v_{B-L}^4}{\rule{0pt}{10pt} 4 |\phi|^4} \right) \right] \,,
\label{eq_coleman-weinberg}
\end{align}
Here, STr denotes the supertrace running over all degrees of freedom of $S_1$ and $S_2$.
$M$ is the corresponding mass matrix, cf.\ Eq.~\eqref{eq_S1S2masses}, and $Q$ an
appropriate renormalization scale, which we have set to $Q^2 = \lambda v_{B-L}^2/2$.


From the resulting scalar potential, $V = V^{F+D}_{\text{SUGRA}} + V_{1\ell}$, we
find the following picture: For $|\phi| > v_{B-L}/\sqrt{2}$, the complex scalars
$s_{1,2}\in S_{1,2}$ are fixed at zero and $\phi$ slowly rolls towards the origin.
At $|\phi| = v_{B-L}/\sqrt{2}$, $\left(m_-^S\right)^2$ becomes negative,
triggering a tachyonic instability.
The Higgs fields acquire a vev and $B$$-$$L$ is broken.
Both the Higgs (which, now that $B$$-$$L$ is broken, is best
parametrized in unitary gauge as a radial degree of freedom
dubbed $\sigma'$ in Ref.~\cite{Buchmuller:2012wn}, cf.\ also
Ref.~\cite{Schmitz:2013gea} for the explicit relation between
$\sigma'$ and the symmetry-breaking Higgs mass eigenstate in
arbitrary gauge) and the inflaton field then quickly fall into
their true vacuum, $|\phi| \rightarrow 0$ and $\sigma' \rightarrow \sqrt{2} \, v_{B-L}$,
eliminating the vacuum energy contributions of the scalar potential
and ending inflation.


\subsection{Slow-roll inflation}


In the slow-roll approximation, the dynamics of the homogeneous inflaton field is governed by
$ 3 H \dot \phi = - \partial V / \partial \phi^* \,$,
where $H$ denotes the Hubble parameter.
For a scalar potential only depending on the absolute value of
$\phi$, cf.\ Eq.~\eqref{eq_coleman-weinberg}, we can rewrite this in
terms of the radial and angular component of
$\phi = \frac{1}{\sqrt{2}} \varphi e^{- i \theta}$,
\begin{align}
 3 H \dot \varphi = - V'(\varphi) \,, \qquad \dot \theta = 0 \,.
\label{eq_slow-roll}
\end{align}
Turning to the quantum fluctuations of the inflaton
field which are visible today in the CMB, we now evaluate the scalar potential and its
derivatives at $\varphi = \varphi_*$, the value of $\varphi$ at  $N_* \approx 55$ e-folds
before the end of inflation, when the reference scale commonly used to describe the CMB
fluctuations left the horizon.
With $\varphi_f$ denoting the value of the inflaton at the end of
inflation,%
\footnote{Here, $\varphi_f$ is determined by either $m_-^S(\varphi_c) = 0$,
cf.\ Eq.~\eqref{eq_S1S2masses}, or by the violation of the slow-roll condition,
i.e.\ $|\eta(\varphi_\eta)|=1$, cf.\ Eq.~\eqref{eq_slow_roll_parameters}, whatever
occurs earlier: $\varphi_f \approx \max\{v_{B-L},
\sqrt{\lambda} M_{\rm Pl}/(\sqrt{8} \pi) \}$.\smallskip}
$\varphi_*$ is given by
\begin{align}
 \varphi_*^2 \approx \varphi_f^2 + \frac{\lambda}{4 \pi^2} M_{\rm Pl}^2 \, N_* 
\label{eq_phistar}
\end{align}
Of particular interest in the following will be the predictions from F-term hybrid
inflation for the amplitude of the scalar fluctuations $A_s$, the scalar spectral
index $n_s$ and the tensor-to-scalar ratio~$r$,
\begin{align}
 \begin{split}
A_s &= \left. \frac{H^2}{8 \pi^2 \epsilon M_{\rm Pl}^2} \right|_{\varphi_*} 
\approx \frac{1}{3} \left( \frac{v_{B-L}}{M_{\rm Pl}} \right)^4 N_* \,,\\
n_s &= \left. 1 - 6 \epsilon + 2 \eta \right|_{\varphi_*} \approx 1 - \frac{1}{N_*}  \,,\\
r &= \left. \frac{A_t}{A_s} = 16 \epsilon \right|_{\varphi_*}
\approx \frac{\lambda}{2 \pi^2} \frac{1}{N_*} \,,
\end{split}
\label{eq_predictions_inflation}
\end{align}
where $A_t = 2 H^2/(\pi^2 M_{\rm Pl}^2)|_{\varphi^*}$ denotes the amplitude of
the tensor fluctuations, and $\epsilon$ and $\eta$  are the so-called slow-roll parameters,
\begin{align}
 \epsilon = \frac{M_{\rm Pl}^2}{2} \left( \frac{V'}{V} \right)^2 \,, \qquad \eta = M_{\rm Pl}^2 \, \frac{V''}{V} \,.
\label{eq_slow_roll_parameters}
\end{align}
Moreover, in Eq.~\eqref{eq_predictions_inflation} we have employed the approximation%
\footnote{Note that, for small values of $\lambda$, the two terms in Eq.~\eqref{eq_phistar}
can be of similar importance, leading to a slight deviation
from the results listed in Eq.~\eqref{eq_predictions_inflation}.}
$\varphi_*^2 \gg \varphi_f^2$.


\begin{figure}
\centering
\includegraphics[width=0.7\textwidth]{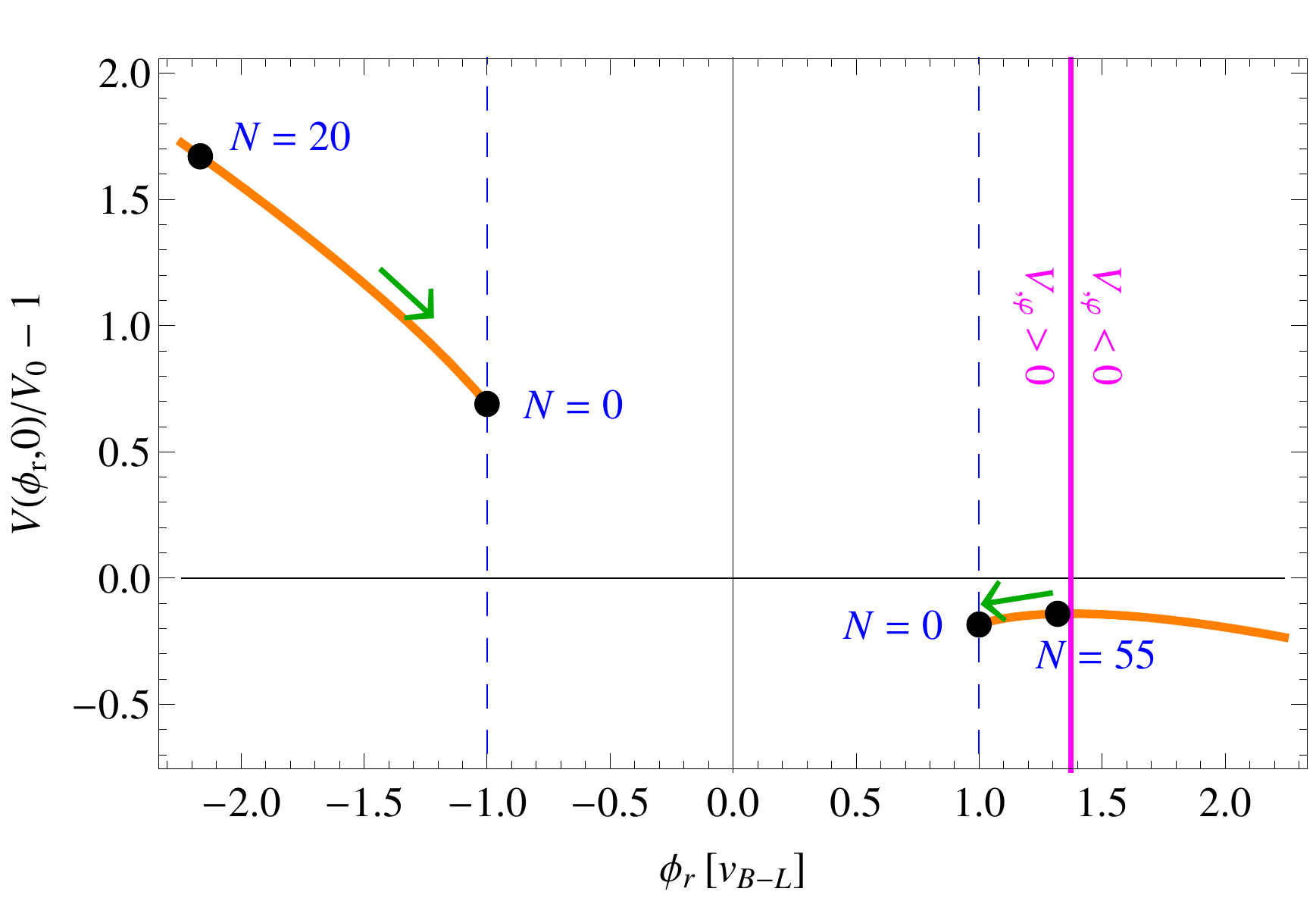}
\caption{Scalar potential for inflation along the real axis in the complex $\phi$ field
space after adding a constant term $W_0$ to the superpotential.
Slow-roll inflation is possible both for $\theta = 0$ and $\theta = \pi$.
Here $\lambda = 4.5 \times 10^{-6}$, $v_{B-L} = 2.9 \times 10^{15}$~GeV and
$m_{\widetilde G} = 47.5$~TeV.}
\label{fig_potential_1D}
\end{figure}


\subsection{F-term hybrid inflation in the light of PLANCK}

\begin{figure}
 \includegraphics[width=1.0\textwidth]{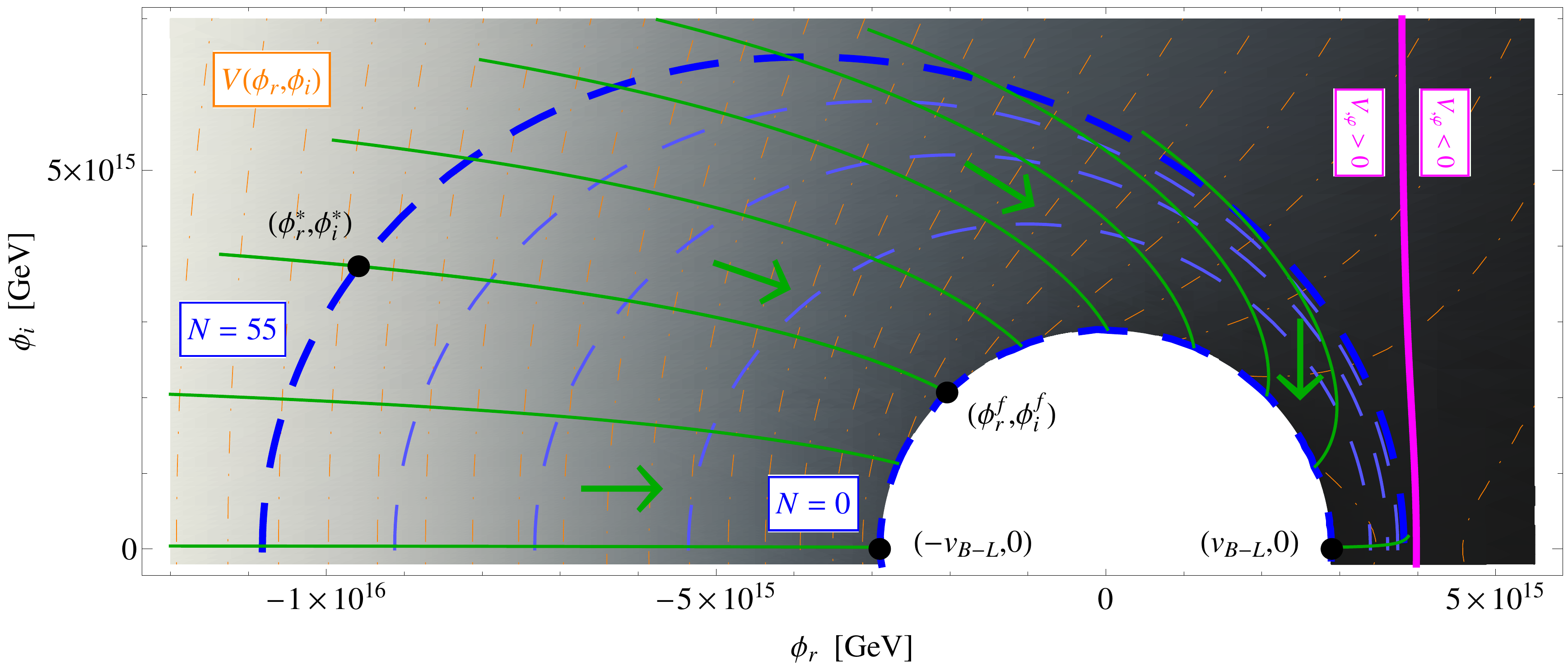}
\caption{Inflationary trajectories in the full two-field inflation model.
Selection of possible trajectories (solid green lines) in the scalar potential
$V(\phi_r, \phi_i)$ depicted by the dot-dashed orange contour lines and the shading.
Lines of constant $N$ are marked by the dashed blue contours, with the beginning
and end of inflation ($N = N_*$ and $N = 0$, respectively) marked by thicker contours.}
\label{fig_multifield_trajectories}
\end{figure}

Comparing these results with the recently published PLANCK data~\cite{Planck:2013jfk, Ade:2013xla},
\begin{align}
 A_s = (2.18 \pm 0.05) \times 10^9\,, \qquad n_s = 0.963 \pm 0.0007 \,, \qquad r < 0.26 \,,
\label{planck_data}
\end{align}
we find that the $B$$-$$L$ breaking scale is fixed to $v_{B-L} \approx 8 \times 10^{15}$~GeV
by requiring the correct normalization of $A_s$, the spectral index $n_s \approx 0.98$ is rather
large and the tensor-to-scalar ratio is easily below the current bound.
In particular, the large value for $n_s$ has raised the question whether F-term hybrid inflation
is still viable in view of the PLANCK results.
To answer this question, we must go beyond the
approximations leading to Eq.~\eqref{eq_predictions_inflation}.
First, we will drop the approximation $\varphi_*^2 \gg \varphi_f^2$,
leading to corrections of the predictions listed in Eq.~\eqref{eq_predictions_inflation}.
Second, taking into account soft supersymmetry breaking, the
superpotential receives a constant term $W_0 = m_{\widetilde G} M_{\rm Pl}^2$
proportional to the gravitino mass~\cite{Buchmuller:2000zm},
leading to an additional contribution to the scalar potential,
studied e.g.\ in Refs~\cite{Nakayama:2010xf,Pallis:2013dxa},
\begin{align}
 V_{m_{\widetilde G}} =  - 2 \sqrt{\lambda} v^2_{B-L} m_{\widetilde G} |\phi| \cos \theta \,.
\label{eq_Vm32}
\end{align}
This term breaks the degeneracy appearing in Eq.~\eqref{eq_coleman-weinberg}, which only
depends on the absolute value  $|\phi|$ of the inflaton field but not on its phase $\theta$.
As a result, the inflationary predictions found in Ref.~\cite{Nakayama:2010xf}
assuming $\theta = \pi$ differ from those in Ref.~\cite{Pallis:2013dxa}, which
uses $\theta = 0$.%
\footnote{Note that Refs.~\cite{Nakayama:2010xf, Pallis:2013dxa} use a different
sign convention in the superpotential, implying $ \theta \rightarrow  \theta + \pi$.}
In particular, for sufficiently large $V_{m_{\widetilde G}}$, we find a hill-top potential
for $\theta = 0$, while for $\theta = \pi$, one still finds a monotonously decreasing potential
along the inflationary trajectory (along the arrows in
Fig.~\ref{fig_multifield_trajectories}), cf.\ Fig.~\ref{fig_potential_1D}.


\begin{figure}
\centering
\includegraphics[width=0.485\textwidth]{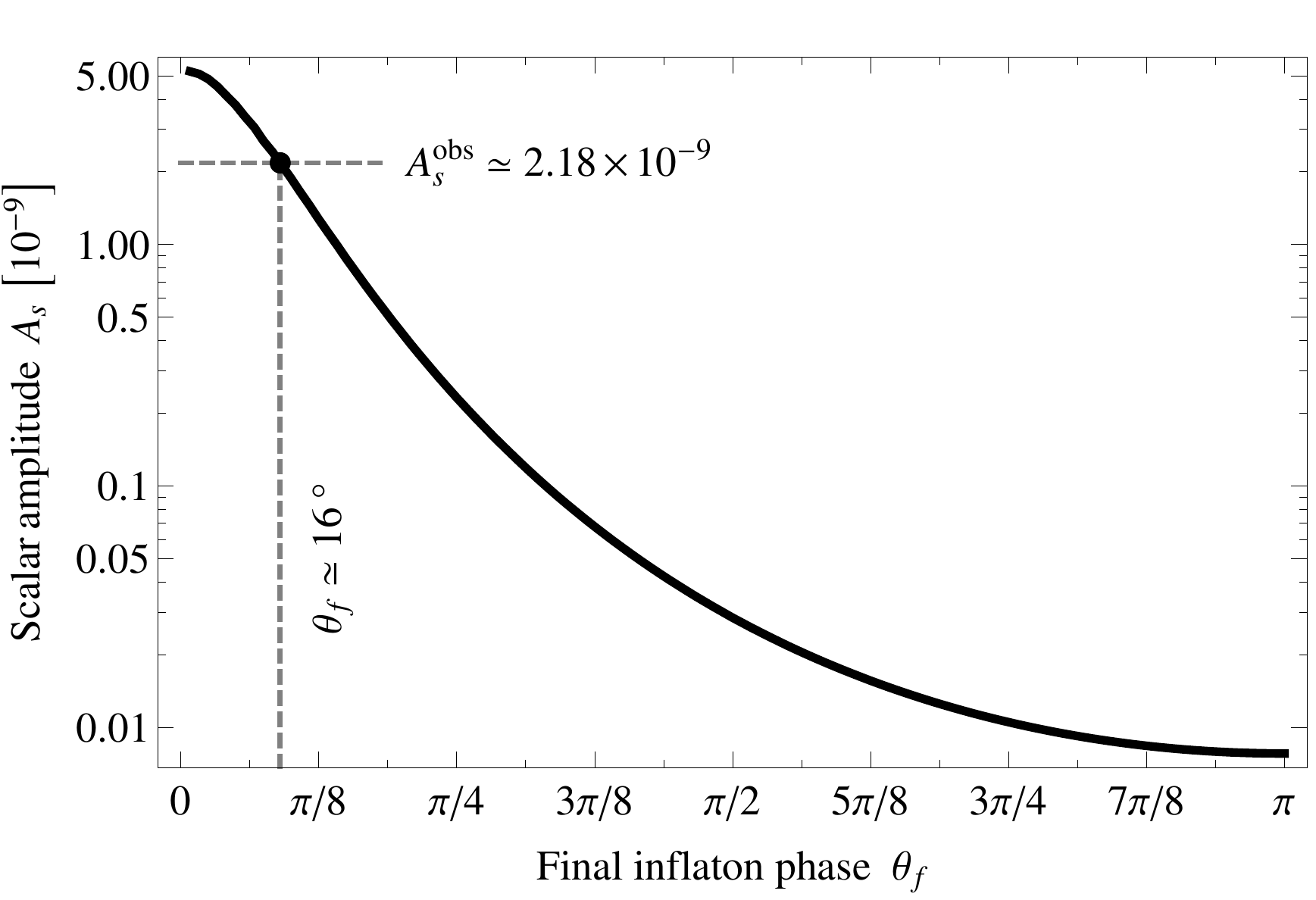}\hfill
\includegraphics[width=0.485\textwidth]{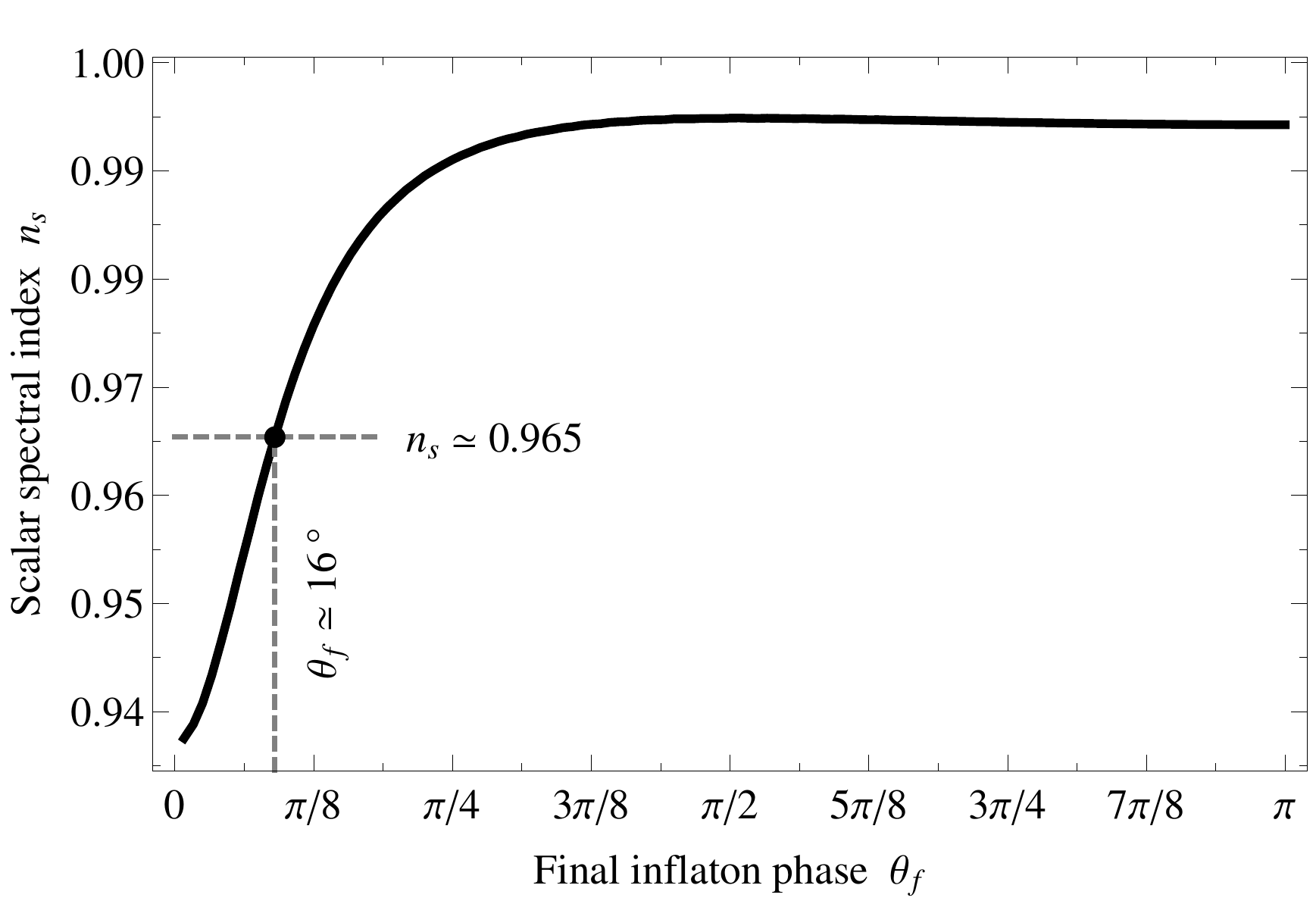}
\caption{Amplitude and spectral index of the scalar primordial fluctuations for
$\lambda = 4.5 \times 10^{-6}$, $v_{B-L} = 2.9 \times 10^{15}$~GeV and
$m_{\widetilde G} = 47.5$~TeV.
The phase of the endpoint of inflation, $\theta_f$, labels different
inflationary trajectories.}
\label{fig_Asns}
\end{figure}

Indeed, these are only two extreme cases for possible inflationary trajectories in
the full two-field inflation model resulting from Eqs.~\eqref{eq_coleman-weinberg}
and \eqref{eq_Vm32}.
In the two-field model, the end point of inflation $\phi_f$ becomes a `critical line'
in the complex $\phi$ plane, with each point on this line representing a possible end
point for inflation.
Hence, in contrast to the single-field case, there is an additional degree of freedom,
i.e.\ the choice of the inflationary trajectory labeled by $\theta_f$.
This is visualized in Fig.~\ref{fig_multifield_trajectories}, which shows a
selection of possible inflationary trajectories (in green) in the scalar potential
(dot-dashed orange contour lines and shading).
Contour lines denoting constant numbers of e-folds $N$ are shown as
dashed blue lines, with the `critical line', $N = 0$, and the onset of
the last $N_*$ e-folds, $N = N_{55}$, emphasized.
To demonstrate the dependence of the model predictions on the choice of the
trajectory, Fig.~\ref{fig_Asns} shows the predictions for the amplitude $A_s$ and
the spectral index $n_s$ as functions of the final phase~$\theta_f$ for
$\lambda = 4.5 \times 10^{-6}$, $v_{B-L} = 2.9 \times 10^{15}$~GeV and
$m_{\widetilde G} = 47.5$~TeV.
For this parameter example, we see that $\theta_f \simeq 16^\circ$ reproduces the
correct amplitude, cf.\ Eq.~\eqref{planck_data}, while simultaneously yielding a
value for the spectral index of $n_s = 0.965$ in very good agreement with the data.


A third possibility of manipulating Eq.~\eqref{eq_predictions_inflation} is by resorting
to a non-minimal K\"ahler potential~\cite{BasteroGil:2006cm}.
This introduces, in particular, a term quadratic in $\phi$ in the scalar potential.
Tuning the expansion coefficients of such a K\"ahler potential, the spectral index
can be tuned to lower values, achieving accordance with the PLANCK data even
for $\theta = \pi$.
However, the quadratic term then comes with a negative sign, implying,
together with the positive $|\phi|^4$ term the existence of a hill-top potential
and a local minimum at $|\phi| \neq 0$ where the inflaton can get trapped.
Avoiding this requires some fine-tuning in the
initial conditions for $|\phi|$~\cite{Nakayama:2010xf}.


On top of that, there are also further constraints which must be taken
into account in a realistic model.
First, the superpotential in Eq.~\eqref{eq_W} will lead to the
production of cosmic strings at the end of inflation, due to the
spontaneous breaking of $U(1)_{B-L}$.
We will come back to this point in Sec.~\eqref{sec_cosmic_strings}.
Moreover, the abundance of nonthermally produced gravitinos,
controlled by the symmetry-breaking scale, the common mass of
the inflaton and $B$$-$$L$ Higgs in the true vacuum, and the
reheating temperature~\cite{Kawasaki:2006gs, Endo:2007ih},
\begin{align}
Y_{\widetilde{G}} \propto \frac{v_{B-L}^2 m_S^2}{T_\text{RH}} \,,
\label{nt-gravitino-abundance}
\end{align}
must be sufficiently low, so that the sum of nonthermal and
thermal, cf.\ Sec.~\ref{sec_reheating}, gravitino abundance does
not produce a gravitino problem~\cite{Weinberg:1982zq, Khlopov:1984pf, Ellis:1984eq,
Ellis:1984er, Moroi:1993mb, Kawasaki:2004yh, Kawasaki:2004qu, Jedamzik:2006xz}.
Note, however, that the nonthermal gravitino abundance can be
suppressed compared to the estimate in Eq.~\eqref{nt-gravitino-abundance},
if the massive particle governing the universe during the reheating phase
decays sufficiently fast.


Taking all of this together, we find that F-term hybrid inflation is indeed
still viable in light of the PLANCK data, however some tuning is required.
Accepting a non-minimal K\"ahler potential with some tuning in its coefficients
as well as in the initial conditions for $|\phi|$, accordance with the PLANCK data
can be achieved for $\theta_f \sim \pi$.
Staying with a minimal K\"ahler potential, one has two options to
reproduce the experimental data.
One possibility is to tune the amplitude of the linear term in Eq.~\eqref{eq_Vm32}
against the CW term in Eq.~\eqref{eq_coleman-weinberg} in the potential,
leading to the situation shown in Fig.~\ref{fig_multifield_trajectories}.
In this case, small values for $n_s$ can be achieved for $\theta_f \sim 0$,
but again tuning of the initial condition for the radial degree $|\phi|$
is necessary to prevent the inflaton from being trapped on the wrong side
of the hill-top potential.%
\footnote{Note that nevertheless an arbitrary amount of e-folds
of inflation can be realized in this setup.}
The second possibility is to allow the linear term to dominate over
the CW term.
Then, however, the initial phase of $\phi$ must be tuned
because otherwise one ends up on a trajectory where inflation does not end,
i.e.\ one risks to `miss' the minimum generated by the CW term at small $|\phi|$,
such that $|\phi|$ always remains larger than the critical value.
For a more detailed analysis of the full two-field inflation model,
cf.\ Ref.~\cite{Buchmuller:2014epa}.


In summary, successful inflation can be achieved, but it imposes constraints on
the $B$$-$$L$ breaking scale and on the coupling $\lambda$.
In the context of the Froggatt-Nielsen flavor model used to
parametrize the Yukawa couplings in Ref.~\cite{Buchmuller:2012wn},
this then constrains the mass of the heaviest of the right-handed
neutrinos $M_1$.
In the following, we will thus consider the restricted parameter space
\begin{align}
v_{B-L} = 5 \times 10^{15}~\text{GeV} \,, \quad
10^9~\text{GeV} \leq M_1 \leq 3 \times 10^{12}~\text{GeV} \,, \quad
10^{-5}~\text{eV} \leq \widetilde{m}_1 \leq 1~\text{eV} \,,
\label{eq_parameter_space}
\end{align} 
where the variation of the effective light neutrino mass parameter
$\widetilde m_1$ accounts for the uncertainties of the Froggatt-Nielsen model.
The values of $v_{B-L}$ and $\lambda$ quoted here correspond to the option
of choosing $\theta_f = \pi$ and using a non-minimal K\"ahler potential,
cf.\ Ref.~\cite{Nakayama:2010xf}.
For a discussion of cosmological $B$$-$$L$ breaking involving smaller
values for $v_{B-L}$, cf.\ Ref.~\cite{Buchmuller:2011mw}.


\section{Tachyonic Preheating and Cosmic Strings}


The end of hybrid inflation induces a negative squared mass term
for the $B$$-$$L$ Higgs field $\sigma'$ in the false vacuum, triggering
the $U(1)_{B-L}$ breaking phase transition.
The cosmological realization of this phase transition is accompanied
by two important nonperturbative processes: tachyonic preheating~\cite{Felder:2000hj}
and the formation of cosmic strings~\cite{Kibble:1976sj}. 


\subsection{Tachyonic preheating}
\label{sec_tachyonic_ph}


\subsubsection*{Phase transition}


Tachyonic preheating is a fast and nonperturbative process triggered by the
tachyonic instability in the scalar potential in the direction of the Higgs field.
As the inflaton field passes a critical value $\phi_c$, the Higgs field $\sigma'$
acquires a negative effective mass squared $-m_\sigma^2$,
with $m_\sigma^2 =  \sqrt{2} \lambda v_{B-L} |\dot \phi_c| t$ in the
linearized equation of motion for $\sigma'$ close to the instability $\phi_c$.
This causes a faster-than-exponential growth of the quantum
fluctuations of the Higgs field $\sigma'_k$ with wave numbers
$|\vec{k}| < m_\sigma $~\cite{Copeland:2002ku}, while the
mean value of the Higgs field remains zero.
Once the amplitude of these fluctuations, $v(t) =  \frac{1}{\sqrt{2}}
\langle \sigma'^2 \rangle = \frac{1}{\sqrt{2}} \langle
\sigma'^2(t, \boldsymbol{ x})\rangle^{1/2}_{\boldsymbol{x}}$,%
\footnote{Here, bold letters indicate 3-vectors.}
reaches $ \langle \sigma'^2 (t_*) \rangle = {\cal O}(v^2_{B-L})$, the curvature
of the potential for the homogeneous background field $\sigma'$ becomes
positive and the usual oscillating behaviour of the modes is
re-established~\cite{Felder:2000hj}, while  $v(t)$ approaches $v_{B-L}$.
A direct consequence of the early phase of exponential growth are high
occupation numbers in the low-momentum Higgs modes and hence a semi-classical
situation with a large abundance of nonrelativistic $B$$-$$L$ Higgs bosons.


A further result of this nonperturbative process is the formation of
`bubble'-like inhomogeneities, which randomly feature different phases
of the complex Higgs field~\cite{GarciaBellido:2002aj,Copeland:2002ku}.
Their initial size is given by the smallest scale amplified during
tachyonic preheating, referred to as $k_*^{-1}$.
These bubbles expand at the speed of light and eventually collide
with each other.
This phase of the preheating process is an important source
of gravitational waves (GWs), cf.~\cite{GarciaBellido:2007dg},
a point to which we will return in Sec.~\ref{sec_GW}.
After this very turbulent phase, the true Higgs vev is reached
in almost the entire volume, with the regimes of false vacuum
reduced to topologically stable cosmic strings, cf.\ Sec.~\ref{sec_cosmic_strings},
separated by the characteristic length scale
$k_*^{-1} \approx (\sqrt{2} \lambda v_{B-L} |\dot \phi_c|)^{-1/3}$.


\subsubsection*{Secondary particle production}

\begin{figure}
\centering
\includegraphics[width=0.485\textwidth]{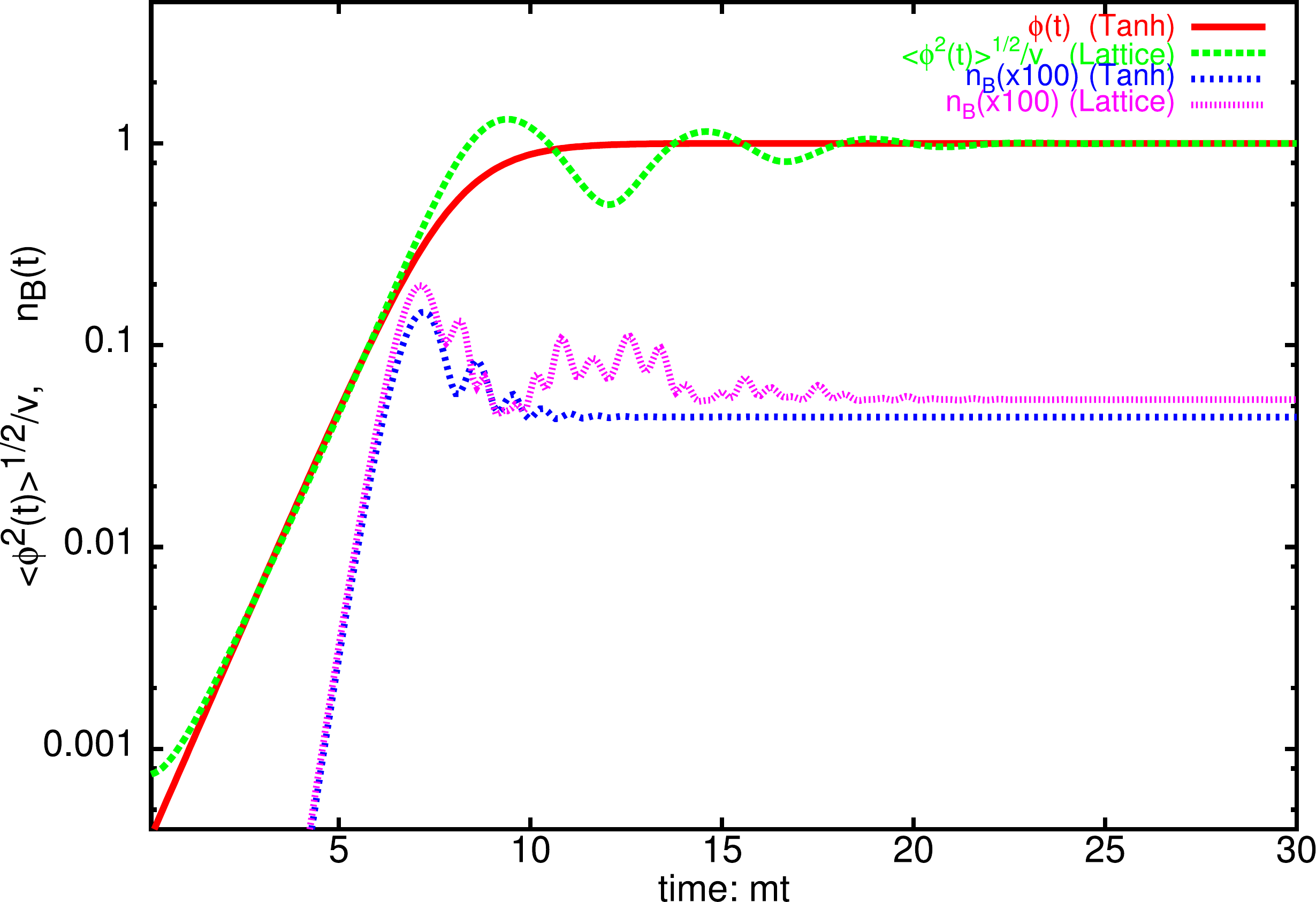}\hfill
\includegraphics[width=0.485\textwidth]{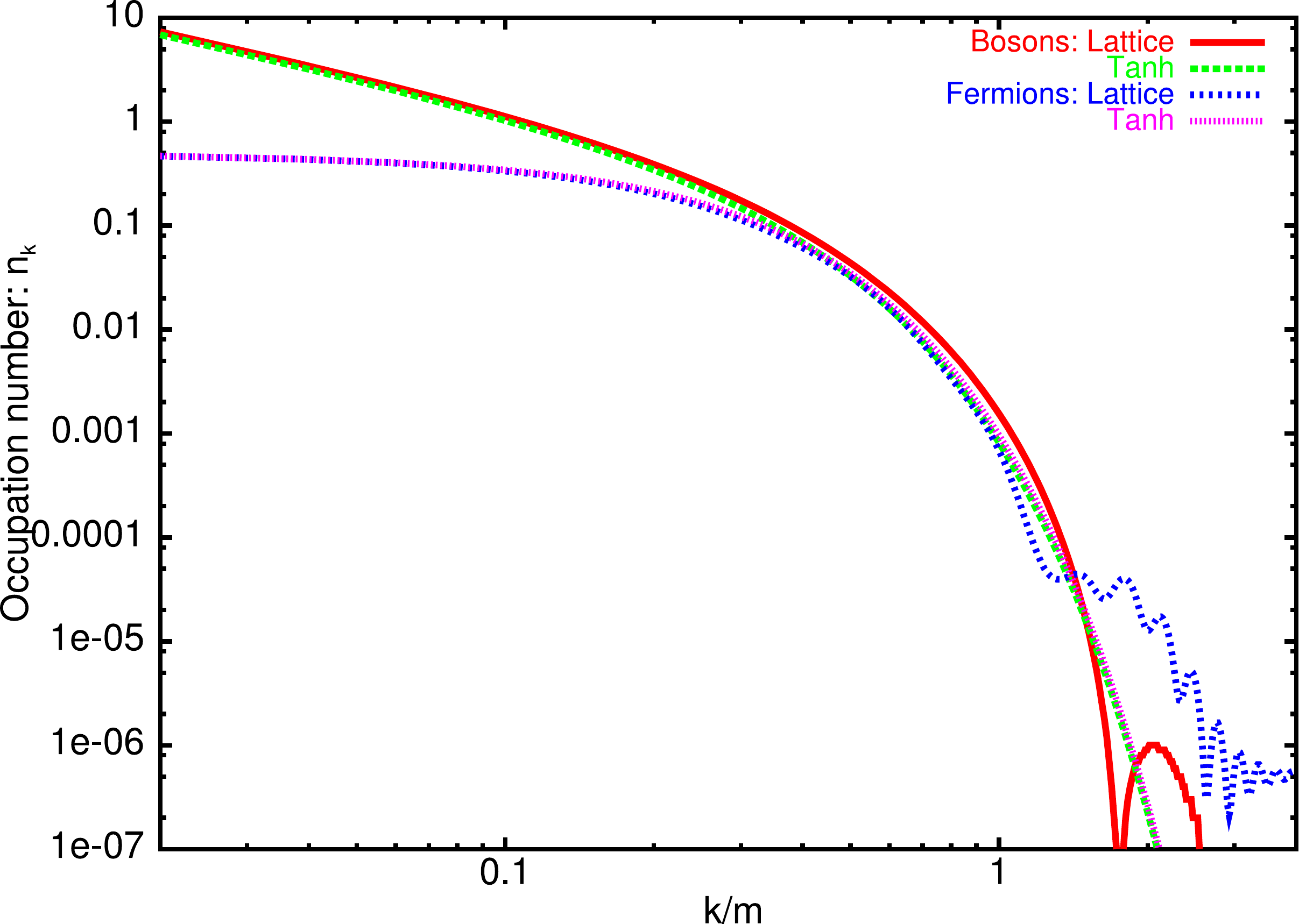}
\caption{Numerical and analytical results for particle production
during tachyonic preheating, taken from Ref.~\cite{GarciaBellido:2001cb}.
\textbf{Left panel}: Evolution of the quantum fluctuations of the Higgs
field $\phi(t) \equiv \sigma'(t)/\sqrt{2}$, normalized to the
symmetry-breaking scale $v \equiv v_{B-L}$, as well as the number
density $n_B$ of bosonic particles coupled to it.
\textbf{Right panel}: Spectrum of occupation numbers for
bosonic and fermionic particles coupled to the Higgs.}
\label{fig_gb}
\end{figure}

The mode equations for the particles coupled to the $B$$-$$L$ Higgs
field, i.e.\ for the gauge, Higgs, inflaton and neutrino supermultiplets,
feature masses proportional to  $v(t)$.
The growth of $\langle \sigma'^2 \rangle$ during tachyonic preheating
thus induces a rapid change of their effective masses.
The resulting particle production was studied in Ref.~\cite{GarciaBellido:2001cb},
with the results depicted in Fig.~\ref{fig_gb}.
Here, for simplicity, an abrupt transition of the inflaton
vev to zero is assumed, introducing the parameter
$m = m_\sigma(\dot \phi_c t \rightarrow  \phi_c)$.%
\footnote{The effect of the  inflaton dynamics on this
nonperturbative particle production requires further investigation.\smallskip}
The left panel shows the evolution of $\langle \sigma'^2 \rangle$ normalized
to the symmetry-breaking scale $v_{B-L}$, calculated using a lattice simulation (green curve).
For comparison, the red curve shows an analytical approximation,
$v(t) = \frac{v_{B-L}}{2}\left(1 + \tanh \frac{m(t- t_*)}{2}\right)$.
The pink and blue curves depict the number densities of bosonic
particles coupled to the Higgs, again calculated using a lattice calculation
and an analytical approximation, respectively.
The right panel examines the momentum distribution of these bosons
(and also of fermions coupled to the Higgs), showing the spectrum of
occupation numbers.
Again, both the numerical and analytical results are shown.
We see that, just like the Higgs bosons themselves, the particles
coupled to it are produced with very low momentum, i.e.\ nonrelativistically.


Based on these results, the energy and number densities for bosons and
fermions coupled to the Higgs boson after tachyonic preheating have been
estimated as~\cite{GarciaBellido:2001cb}%
\footnote{These estimates can be significantly
enhanced by quantum effects \cite{Berges:2010zv}, which also require further investigation.}
\begin{align}
\begin{split}
&\rho_B/\rho_0 \simeq 2 \times 10^{-3} \, g_\sigma \, \lambda \,
f(x_1, 1.3) \,, \phantom{.5} \qquad  n_B(x_1) \simeq 1 \times 10^{-3} \, 
g_\sigma  \, m_S^3 \, f(x_1, 1.3)/x_1 \,,\\
&\rho_F/\rho_0 \simeq 1.5 \times 10^{-3} \, g_\sigma \, \lambda \,
f(x_1, 0.8) \,, \qquad  n_F(x_1) \simeq 3.6 \times 10^{-4} \,  g_\sigma \, m_S^3 \,
f(x_1, 0.8)/x_1 \,, \label{eq_partprod}
\end{split}
\end{align}
with $f(x_1, x_2) = (x_1^2 + x_2^2)^{1/2} - x_2$ and $x_1 = m_i/m_S$,
where $m_i$ denotes the mass of the respective particle in the true vacuum
and $g_\sigma$ counts its spin and internal degrees of freedom.


\subsection{Cosmic strings}
\label{sec_cosmic_strings}


Due to the non-trivial topology of its vacuum manifold, the Abelian Higgs
model underlying the $B$$-$$L$ phase transition gives rise to solitonic
field configurations, so-called cosmic strings; for reviews, cf.\
e.g.~Refs.~\cite{Vilenkintextbook, Hindmarsh:2011qj, Lenz:2001me}.
These cosmic strings are formed during the process of tachyonic preheating
and are topologically stable.
The evolution of the resulting network is governed by the intersection
of the infinite strings, which leads to the formation of closed loops
separated from the infinite string, as well as by the energy loss due
to the emission of GWs, Higgs and gauge particles.
After a relaxation time, the network reaches the scaling regime, i.e.\ the
typical length scale of the cosmic string network remains constant relative
to the size of the horizon.
This implies that a constant fraction of the total energy density is
stored in cosmic strings throughout the further evolution of the universe
and that there are ${\cal O}(1)$ cosmic strings per 
Hubble volume. 


In the scaling regime, the cosmic string network is characterized
by the energy per unit length~$\mu$.
In the Abelian Higgs model, which is based on a field theory
featuring a spontaneously broken local $U(1)$ symmetry, $\mu$ is
given by~\cite{Hill:1987ye}
\begin{align}
\label{eq_mu}
\mu = 2 \pi B(\beta) v_{B-L}^2 \,,
\end{align}
where $\beta = (m_S / m_G)^2 = \lambda/(8 g^2)$ is the ratio of
the masses of the symmetry-breaking Higgs boson and the gauge boson
in the true vacuum, and $B(\beta)$ is a slowly varying function
parameterizing the deviation from the Bogomol'nyi bound,
\begin{align}
B(\beta) \simeq \left\{\begin{array}{ll} 
1.04 \, \beta^{0.195} \,, & \mbox{if } 10^{-2} < \beta \ll 1 \\ 
2.4 / \ln\left(2/\beta\right) \,, & \mbox{if } \beta < 10^{-2}  \end{array}\right. \,.
\label{eq_B}
\end{align}
For the special case of $\beta = 1$ the Bogomol'nyi bound
is saturated and $B(1) = 1$~\cite{deVega:1976mi}. 


Further important quantities describing the string network
are the cosmic string width, given by $m_G^{-1}$ in the Abelian Higgs
model, and the length scale $\xi$ separating two strings.
From Sec.~\ref{sec_tachyonic_ph}, we know that the characteristic
length separating two strings at the time of their formation is
\begin{align}
\label{eq_xi}
\xi \simeq k_*^{-1} = ( \sqrt{2} \lambda v_{B-L} |\dot{\phi}_c|)^{-1/3} \,.
\end{align}
This also determines the relaxation time of the cosmic
string network, $\tau_{\text{string}} \sim \xi$~\cite{Copeland:2002ku, Hindmarsh:2008dw}.
Note that in the Nambu-Goto model, an alternative to the Abelian
Higgs cosmic string model which assumes infinitely thin cosmic strings,
the energy scale $\mu$ is an input parameter.


\subsubsection*{Observational prospects}


So far, no experimental evidence for the existence of cosmic strings has been found.
However, current and upcoming experiments are starting to seriously probe
the cosmologically interesting regions of the parameter space. 
First, cosmic strings give rise to anisotropies in the CMB temperature map.
They  distort the surface of last scattering of the CMB photons, leaving an
imprint on the spectrum observable today.
Since the CMB photons observable today stem from roughly $10^5$ Hubble
patches during recombination, these observations are mainly sensitive
to the effect of long (Hubble-sized) strings at recombination and not
to small cosmic string loops.
In contrast to the perturbations due to inflation, these anisotropies
are not phase-correlated across distant Hubble patches and hence the
resulting multipole spectrum of the two-point correlation function is
suppressed at large scales.
Moreover, whereas the primordial power spectrum due to inflation is
(nearly) scale-invariant, the anisotropies on the last scattering surface
due to cosmic strings are governed by a characteristic scale.
The resulting spectrum thus features a single broad peak 
associated with this scale.
Due to the re-scattering of a fraction of the CMB photons at
reionisation, the CMB spectrum is, to a lesser extent, also sensitive
to the long cosmic strings present at reionisation.
This leads to a second, smaller peak in the spectrum, in 
particular visible in the power spectrum of the B-mode polarization,
cf.\ e.g.~Ref.~\cite{Kuroyanagi:2012jf} for a recent analysis.
The fraction of the amplitude of the scalar power spectrum due
to a possible cosmic string contribution to the CMB temperature
anisotropies is conventionally measured at the $\ell=10$ multipole
and is referred to as $f_{10}$.
The PLANCK data implies that $f_{10}$ can at most
be a few percent, $f_{10} < 2.8 \%$~\cite{Ade:2013xla}.


Second, the gravitational field of cosmic strings gives rise to
weak and strong lensing effects of (CMB) photons on their way from
the surface of last scattering or from an astrophysical source to us.
The non-observation of such effects puts a bound on the string tension $\mu$.
Again, this effect is mainly sensitive to long (Hubble-sized) strings.
Third, the energy emitted by cosmic strings in the scaling regime is
at least partly emitted in form of GWs.
Due to their extremely weak coupling, these can then propagate
freely through the universe and are therefore, in principle,
detectable today.
We will come back to the resulting GW background and the discovery
potential of current and upcoming GW experiments in detail in Sec.~\ref{sec_GW}.
Finally, the Abelian Higgs cosmic string model entails the
emission of massive radiation from cosmic strings, i.e.\ the emission
of the Higgs and gauge particles whose field configurations form the string.
If this mechanism is still active at late times, it could yield
ultra-high-energetic cosmic rays and GeV-scale $\gamma$-rays, which
have not been observed.
This, too, can be translated into a (model-dependent) bound
on $\mu$~\cite{Sigl:1995kk,Protheroe:1996pd,Bhattacharjee:1997in,Sigl:1998vz,Wichoski:1998kh}.
Currently the most stringent and model-independent bound
on the cosmic string tension comes from CMB observations,
$G \mu < 3.2 \times 10^{-7}$~\cite{Ade:2013xla}, and we shall
mainly employ this bound in the following.


\subsubsection*{Numerical simulations and theoretical uncertainties}


A quantitative understanding of the formation of cosmic strings, the dynamics
of the cosmic string network and the energy loss mechanism during the scaling
regime requires lattice simulations.
Performing these is extremely challenging due to the huge range
of scales involved in the problem~\cite{Hindmarsh:2011qj}:
the width of the string remains constant while the scales of the
network are blown up as the universe expands.
Or, in comoving coordinates, the comoving width of the string shrinks,
until it becomes comparable with the lattice spacing and the simulation
loses its validity.
There have been different approaches to tackle this problem.
Simulations based on solving the field equations for the Abelian
Higgs (AH) model set the comoving width to a finite constant before
it comes too close to the lattice spacing~\cite{Vincent:1997cx, Moore:2001px, Hindmarsh:2008dw}.
Simulations based on the Nambu-Goto (NG) string model assume cosmic
strings to be infinitely thin, i.e.\ strictly one-dimensional objects, 
throughout the simulation~\cite{Albrecht:1989mk, Vanchurin:2005pa, Olum:2006ix, BlancoPillado:2011dq}.
The outcome of simulations based on these two models is dramatically different.
The AH simulations show the formation of large, Hubble-sized structures
which lose their energy predominantly by emitting massive radiation,
i.e.\ particles of the Higgs and gauge fields forming the string configuration.
The NG simulations on the other hand display the formation of small loops,
which lose their energy into GWs.
The size of these loops is thought to be controlled by gravitational
backreaction, but is as yet undetermined~\cite{Hindmarsh:2011qj}.
Concerning the network of long strings, both simulations, however,
yield a similar result~\cite{ Hindmarsh:2011qj}.
Which of these two simulations methods is closer to reality is
currently an open question. 


In the following, we will adopt the following hypothesis:
For early times, while the comoving cosmic string width is large
compared to the lattice spacing, the AH simulation describes the $U(1)$ phase
transition very well.
We will thus use the results from these simulations when
discussing the formation and early evolution of cosmic strings.
For late times, the AH simulations become questionable and the NG
approximations of infinitely thin strings appears reasonable.
Hence, for late times, in particular when discussing possible GW signatures
from cosmic strings, cf.~Sec.~\ref{sec_GW}, we shall discuss
both the AH as well as the NG results.


\section{Reheating}
\label{sec_reheating}


Tachyonic preheating nonperturbatively generates a large abundance of
nonrelativistic $B$$-$$L$ Higgs bosons as well as, to a much lesser extent,
nonrelativistic abundances of the particles coupled to the Higgs boson,
cf.\ Sec.~\ref{sec_tachyonic_ph}.
Among these are the particles of the $B$$-$$L$ gauge supermultiplet,
which decay quickly due to their comparatively strong gauge interactions.
This sets the initial conditions for the following slow, perturbative
reheating process, depicted by the solid arrows in the left panel of
Fig.~\ref{fig:overview_numdens}:
The particles from the symmetry-breaking sector decay into
particles of the $N_1$ supermultiplet.
These (s)neutrinos, just as the (s)neutrinos produced through
gauge particle decays and tachyonic preheating as well as thermally
produced (s)neutrinos, decay into MSSM particles, thereby
generating the entropy of the thermal bath as well as a
lepton asymmetry~\cite{Buchmuller:2005eh}.
Finally, the thermal bath produces a thermal abundance of gravitinos,
which will turn out to be in the right ball-park to account for the
observed relic density of dark matter. 

The main tool to obtain a time-resolved description of this
reheating process are Boltzmann equations, which describe the
evolution of the phase space densities of the various particles
species due to decay and scattering processes in an expanding universe.
After briefly introducing the formalism of Boltzmann
equations in Sec.~\ref{subsec_boltzmann_formalism}, we will
turn to the implications for leptogenesis and dark matter
production in Sec.~\ref{subsec_reheating_pheno}.
The results presented here are based on the analyses of
Refs.~\cite{Buchmuller:2010yy, Buchmuller:2011mw,Buchmuller:2012wn}. 


\subsection{Boltzmann equations}
\label{subsec_boltzmann_formalism}


\begin{figure}
\centering
\includegraphics[width=0.26\textwidth]{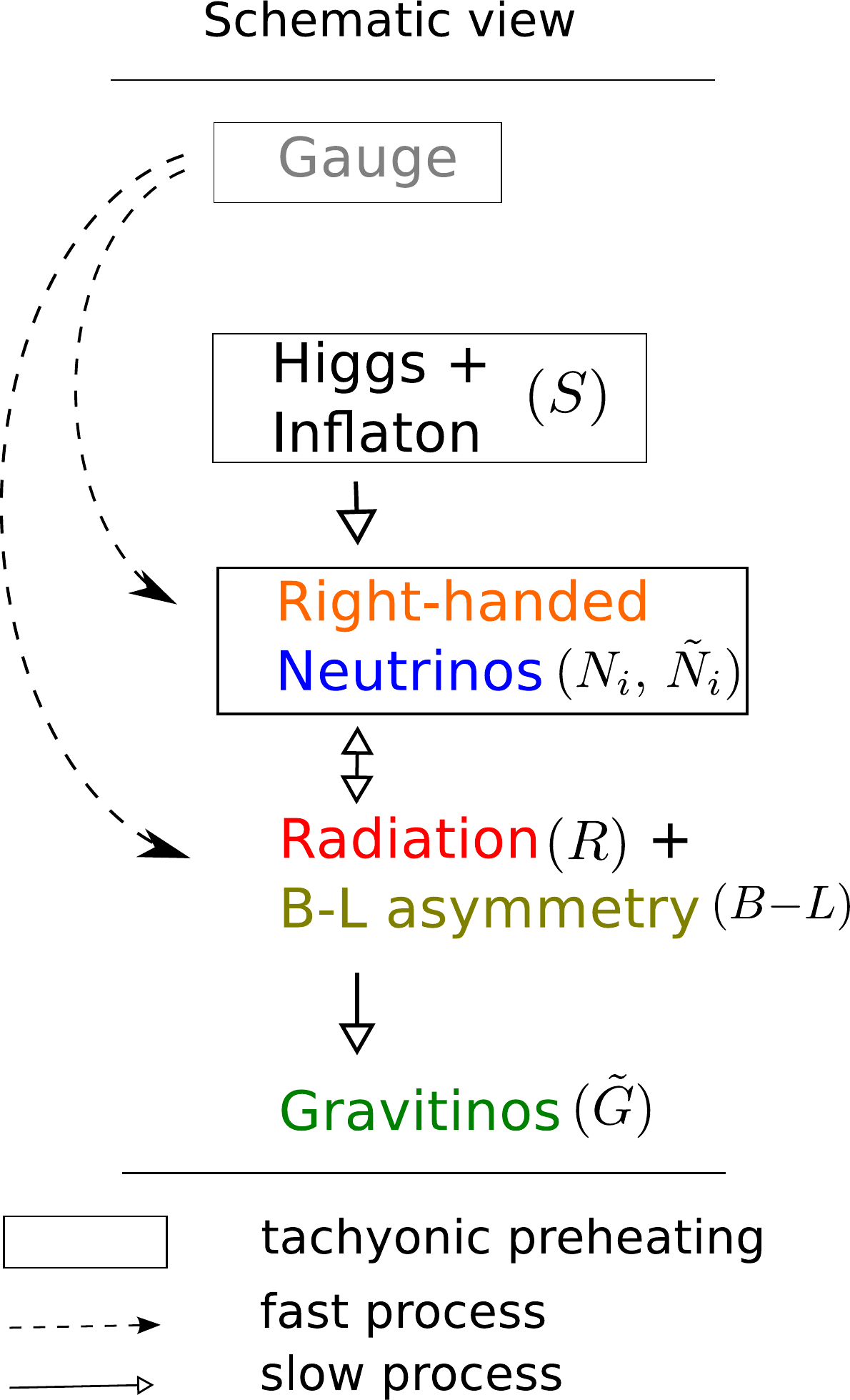}\hfill
\includegraphics[width=0.65\textwidth]{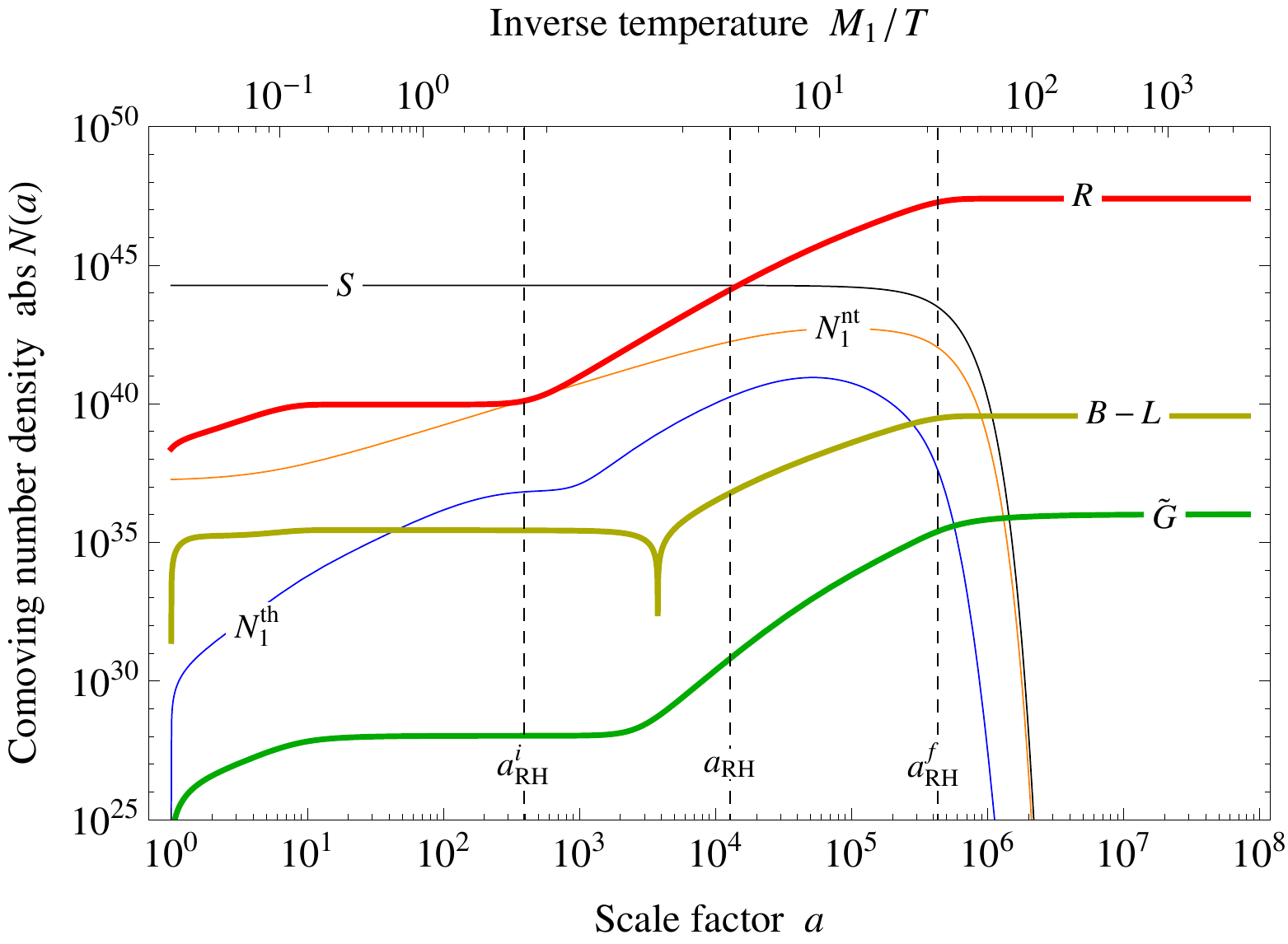}
\caption{Evolution of the comoving number densities
during the reheating process. \textbf{Left panel}:
schematic overview, distinguishing production via tachyonic
preheating, fast decay processes of the $B$$-$$L$ gauge
sector and slow processes described by Boltzmann equations. 
\textbf{Right panel}: Comoving number densities of
the particles of the $B$$-$$L$ Higgs sector ($S$), the
thermal and nonthermal (s)neutrinos $(N_1^\text{th}, \, N_1^\text{nt})$,
the MSSM radiation ($R$), the gravitinos $(\widetilde G)$ and
the $B$$-$$L$ asymmetry ($B$$-$$L$).
Obtained by solving the Boltzmann equations for
$v_{B-L} = 5 \times 10^{15}$~GeV, $M_1 = 5.4 \times 10^{10}$~GeV,
$\widetilde m_1 = 4.0 \times 10^{-2}$~eV, $m_{\widetilde G} = 100$~GeV
and $m_{\tilde g} = 1$~TeV. From Ref.~\cite{Buchmuller:2012wn}.}
\label{fig:overview_numdens}
\end{figure}


The evolution of the phase space density $f_{X}(t,p)$ of
a particle species $X$ is determined by a coupled set
of Boltzmann equations,
\begin{align}
\label{eq_boltzmann}
E\left( \frac{\partial}{\partial t} - H p \frac{\partial}{\partial p} \right)
f_{X}(t,p) = \sum_{i'j'..} \sum_{ij..} C_X(Xi'j'.. \leftrightarrow ij..) \,,
\end{align}
augmented by the Friedmann equation, which governs
the evolution of the scale factor.
The left-hand side of Eq.~\eqref{eq_boltzmann} describes
the evolution of the phase space density in an expanding
Friedman-Robertson-Walker (FRW) universe, whereas the
collision operators $C_X$ on the right-hand side account
for all relevant scattering, decay and inverse decay processes
that the particle $X$ is involved in.
The set of Boltzmann equations we have to solve here
is determined by the allowed interactions of the underlying
particle physics model, cf.\ the solid arrows in
the left panel of Fig.~\ref{fig:overview_numdens}.


From the phase space density $f_X(t,p)$ one directly obtains
the comoving number density $N_X(t)$, i.e.\ the number of $X$
particles in a volume $(a/\text{GeV})^3$, and the energy
density $\rho_X(t)$ by integrating over momentum space,
\begin{align}
\begin{split}
N_X(t) &  = \left(\frac{a(t)}{\text{GeV}} \right)^3\, n_X  = 
\left(\frac{a(t)}{ \text{GeV}} \right)^3\, g_X \int \frac{d^3p}{(2 \pi)^{3}} \, f_X(t,p) \,, \\
\rho_X(t) & = g_{X} \int \frac{d^3p}{(2 \pi)^3} \, E_{X}(p) \, f_X(t,p) \,,
\end{split}
\end{align}
with $a$ denoting the scale factor.
A rescaling of $a$ leaves the physical number density
$n_X$ invariant.
For convenience, we will thus set $a_{\text{PH}} \equiv 1$
at the end of preheating. 
In the following, decay rates $\Gamma$, comoving number
densities $N$ and energy densities $\rho$ will sometimes
appear with upper and lower indices.
In this case, the lower index refers to the
particle species under consideration, while the upper
index refers to its origin, e.g.\ its parent particle
or `PH' for preheating.


\subsection{Outcome of the reheating process \label{subsec_reheating_pheno}}


Solving the Boltzmann equations with the initial conditions
given by tachyonic preheating and the successive decay of the
$B$$-$$L$ gauge fields yields a time-resolved picture of the
evolution of all particle species.
In the right panel of Fig.~\ref{fig:overview_numdens}, we
show an overview of the resulting comoving number densities
for a representative parameter point.


\subsubsection*{A two-stage reheating process}


After the end of preheating, the lion's share of the energy is
stored in nonrelativistic $B$$-$$L$ Higgs bosons.
Assuming a hierarchical spectrum of heavy Majorana neutrinos,
these decay exclusively into heavy, typically relativistic
(s)neutrinos of the first generation,
thereby forming the main part of the right-handed (s)neutrino population.
The decay of these (s)neutrinos then generates a thermal bath of MSSM particles.
The process of reheating is hence governed by the interplay of two time scales,
the vacuum decay rate of the nonrelativistic Higgs bosons $\Gamma_S^0$
and the effective decay rate of the neutrinos produced in the Higgs
boson decays $\Gamma_{N_1}^S$.
The latter differs from the zero-temperature decay rate
$\Gamma_{N_1}^0$ due to the time dilatation of the relativistic neutrinos,
\begin{align}
\begin{split}
\Gamma_S^0 & = \frac{1}{32 \pi} \left( \frac{M_1}{v_{B-L}} \right)^2
m_S \left(1 - 4 \frac{M_1^2}{m_S^2} \right)^{1/2} \,, \\
\Gamma_{N_1}^S &  := \Gamma_{N_1}^S (a_\text{RH}) = \gamma^{-1}(a_\text{RH})
\, \Gamma_{N_1}^0\, \quad \text{   with} \quad \gamma^{-1}(a) =
\left \langle \frac{M_1}{E_{N_1}} \right \rangle_a^{(S)} \,, \quad
\Gamma_{N_1}^0 = \frac{1}{4 \pi} \frac{\widetilde m_1 M_1^2}{v^2_{\text{EW}}} \,.
\end{split}
\end{align}
In most of the viable parameter space, we find $\Gamma_S^0 < \Gamma_{N_1}^S$.
In this case, $\Gamma_S^0$ determines the overall time scale of the
reheating process.
On the contrary, $a_\text{RH}$, defined by $H(a_\text{RH}) = \Gamma_{N_1}^S(a_\text{RH})$,
marks a characteristic point in the middle of the reheating process,
which will be particularly relevant for determining the reheating temperature.
Once the Higgs bosons decay into neutrinos, these decay nearly
instantaneously into MSSM particles, so that the era of Higgs domination
is directly followed by the radiation dominated epoch.
On the other hand, if $\Gamma_S^0 > \Gamma_{N_1}^S$, the effective
neutrino decay rate governs the time scale of reheating.
The energy density is then successively dominated by nonrelativistic Higgs bosons,
relativistic nonthermal neutrinos and finally relativistic MSSM particles
in thermal equilibrium.


\begin{figure}
\centering
\includegraphics[width=0.7\textwidth]{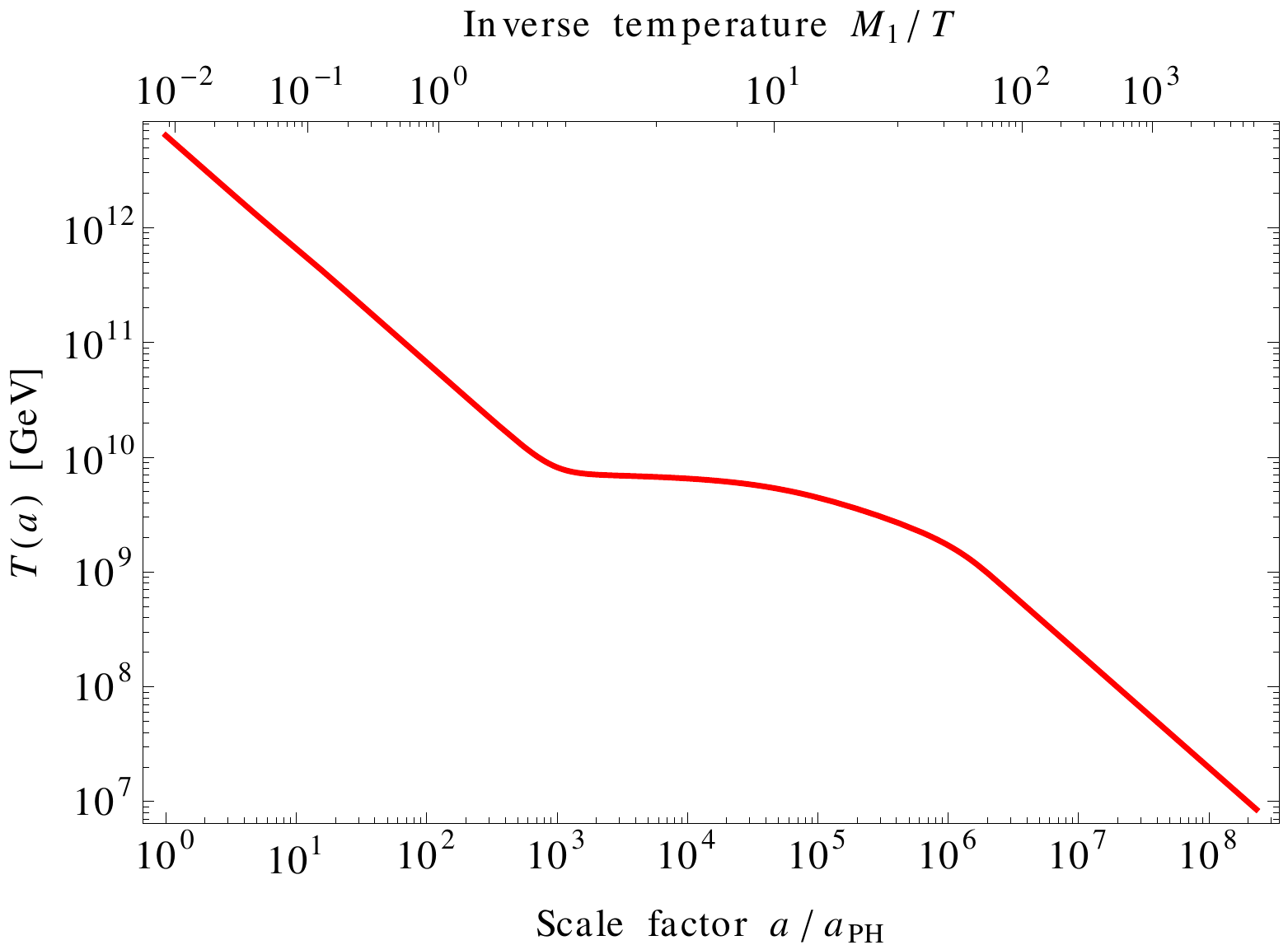}
\caption{\rm Temperature of the thermal bath for
the same parameter values as in Fig.~\ref{fig:overview_numdens}.
From Ref.~\cite{Buchmuller:2012wn}.}
\label{fig:temperature}
\end{figure}


Solving the Boltzmann equations allows us to determine
the temperature of the thermal bath throughout the reheating process.
As a consequence, the `reheating temperature' is no longer a
cosmological input parameter, but is rather determined by the
parameters of the $B$$-$$L$ Higgs and neutrino sector.
In Fig.~\ref{fig:temperature}, we show the resulting evolution
of the temperature as a function of the scale factor.
A remarkable feature is the epoch of nearly constant
temperature during the main part of the reheating process,
which arises because the entropy production in neutrino
decays just compensates the expansion of the universe.
A typical value for this plateau is given by
$T_\text{RH}^N \equiv T(a_\text{RH})$, with $a_\text{RH}$
as defined above.
The dashed vertical lines labeled $a_\text{RH}^i$ and
$a_\text{RH}^f$ in Figs.~\ref{fig:overview_numdens} and
\ref{fig:temperature} mark the beginning and the
end of the reheating process, defined as the period
when the effective production rate of MSSM particles
exceeds the Hubble rate.


\subsubsection*{Thermal and nonthermal leptogenesis}

\begin{figure}
\centering
\includegraphics[width=0.7\textwidth]{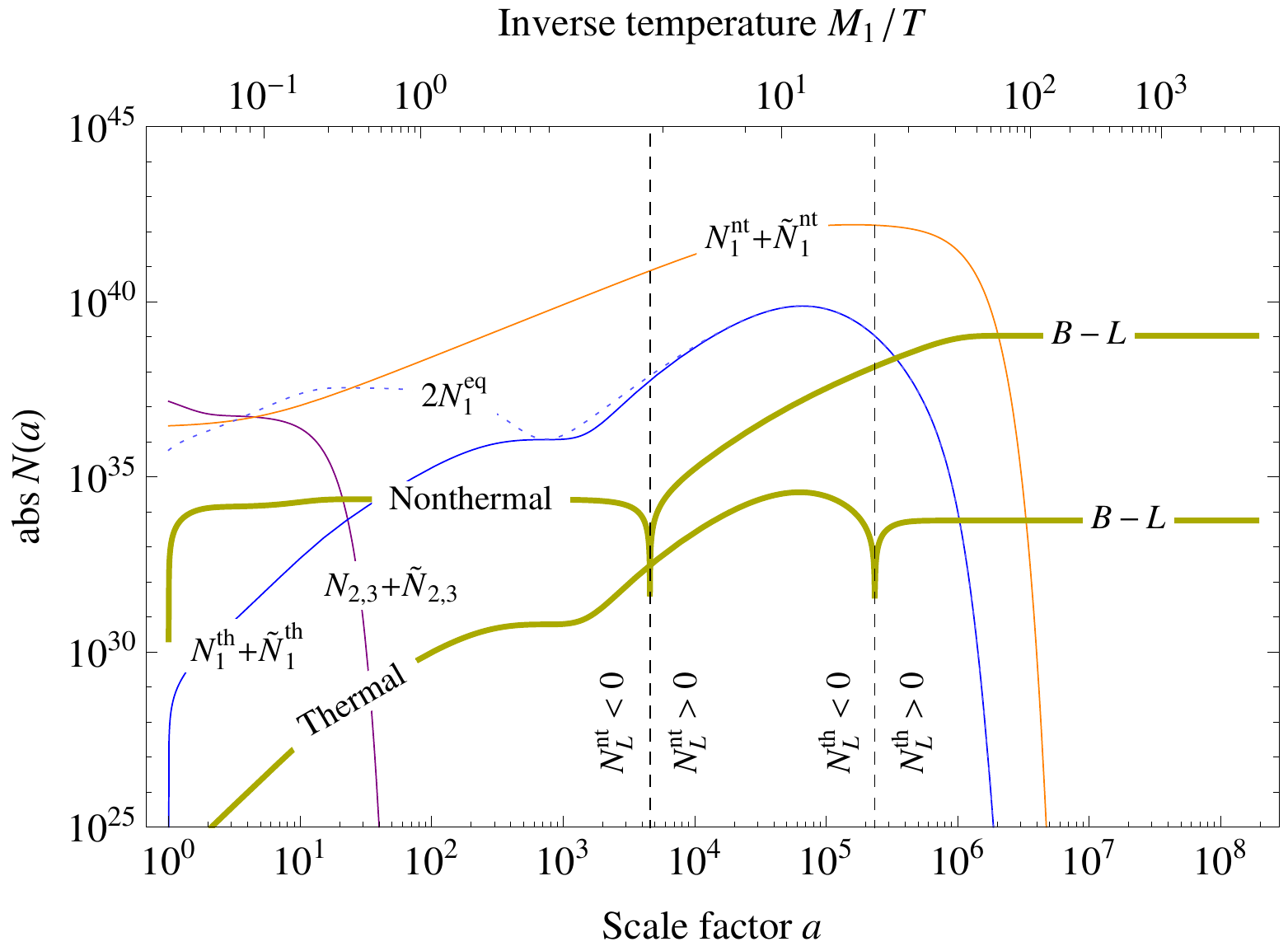}
\caption{Comoving number densities for the nonthermal
($N_L^\textrm{nt}$) and thermal ($N_L^\textrm{th}$) contributions
to the total lepton asymmetry as well as all (s)neutrino species
($N_{1}^{\textrm{nt}} \! + \! \tilde{N}_{1}^{\textrm{nt}}$,
$N_{1}^{\textrm{th}} \! + \! \tilde{N}_{1}^{\textrm{th}}$,
$2 N_1^\textrm{eq}$ for comparison and 
$N_{2,3} \! + \! \tilde{N}_{2,3}$) as functions of the scale factor $a$.
The vertical lines mark the changes in the signs of the
two components of the lepton asymmetry.
From Ref.~\cite{Buchmuller:2012wn}.
\label{fig:BL_asymmetry}}
\end{figure}

The decays of the thermally and nonthermally produced
neutrinos give rise to a thermal and a nonthermal $B$$-$$L$
asymmetry, as depicted in Fig.~\ref{fig:BL_asymmetry}.


The nonthermal lepton asymmetry receives a first
contribution from the decay of the heavy (s)neutrinos
of the second and third generation.
To clearly distinguish this contribution from the main
contribution arising due to the decay of the first-generation
(s)neutrinos, we have assigned opposite signs to the
parameters $\epsilon_{2,3}$ and $\epsilon_1$ quantifying
the $CP$ asymmetry in decays of the respective neutrino generations.
This entails the change of sign visible at $a \simeq 4.6 \times 10^3$
in Fig.~\ref{fig:BL_asymmetry}, when the decay of $N_1^S$ neutrinos
becomes efficient and the main part of the nonthermal asymmetry is produced.
washout effects are negligibly small throughout this process and hence,
once the production of the nonthermal asymmetry becomes inefficient,
the asymmetry freezes out.


The production of the thermal asymmetry is driven by the
deviation of the thermal (s)neutrino abundance from the equilibrium value.
This leads to an initially negative asymmetry with a rapidly increasing
absolute value.
This increase slows down as the thermal (s)neutrino abundance
approaches the equilibrium value.
At around $a \simeq 6.3 \times 10^4$, washout processes
start to play a role, leading to a decrease of the asymmetry.
The situation rapidly changes when the thermal (s)neutrino
abundance overshoots the equilibrium abundance towards the
end of the reheating process.
This generates an asymmetry with an opposite sign, which
overcompensates the asymmetry generated so far.
Shortly after, both the washout rate and production rate
drop significantly below the Hubble rate and the asymmetry
freezes out.


The final values of thermal and nonthermal asymmetry as depicted
in Fig.~\ref{fig:BL_asymmetry} allow us to infer the present
baryon asymmetry $\eta_B$ as well as its composition
in terms of a nonthermal ($\eta_B^{\textrm{nt}}$) and
a thermal ($\eta_B^{\textrm{th}}$) contribution,
\begin{align}
\eta_B = \frac{n_B^0}{n_\gamma^0} = \eta_B^{\textrm{nt}} + \eta_B^{\textrm{th}} \,,\quad
\eta_B^{\textrm{nt},\textrm{th}} =  C_{\textrm{sph}} \frac{g_{*,s}^0}{g_{*,s}^\text{RH}}
\left.\frac{N_{L}^{\textrm{nt},\textrm{th}}}{N_\gamma}\right|_{a_f} \,.
\label{eq:etaBntth}
\end{align}
Here, $C_{\textrm{sph}} = 8/23$ is the sphaleron conversion factor,
$g^\text{RH}_{*,s} = 915/4$ and $g_{*,s}^0 = 43/11$ are the effective
numbers of relativistic degrees of freedom in the MSSM that enter the
entropy density $s$ in the high- and
low-temperature regime, respectively,
$N_L^\text{nt,th}$ refers to the comoving number densities of the nonthermal
and thermal contributions to the lepton asymmetry, and
$N_\gamma = g_\gamma / g_{*,n} \, N_{r}$ is the comoving number density of photons.
For our parameter example we find
\begin{align}
\eta_B \simeq 3.7 \times 10^{-9} \,, \quad
\eta_B^{\textrm{nt}} \simeq 3.7 \times 10^{-9} \,, \quad
\eta_B^{\textrm{th}} \simeq 1.9 \times 10^{-14} \,.
\label{eq:etaBres}
\end{align}
Note that, to obtain these values, we have set the $CP$ violation parameter
in the first generation neutrino decays $\epsilon_1$ to the maximally
allowed value, cf.\ Ref.~\cite{Buchmuller:2012wn}.
Hence $\eta_B$ in Eq.~\eqref{eq:etaBres} yields an upper bound on the
baryon asymmetry produced in this setup and is thus perfectly compatible
with the observed value,
$\eta_B^{\textrm{obs}} \simeq 6.2 \times 10^{-10}$~\cite{Komatsu:2010fb}.
In fact, the Froggatt-Nielsen flavor model employed in our numerical analysis typically
predicts a value for $\epsilon_1$ that is smaller than the maximally possible
value by roughly a factor of $\mathcal{O}(10)$,
cf.\ Ref.~\cite{Buchmuller:2011tm}, implying excellent agreement
between prediction and observation for this parameter
example, $\eta_B \simeq \eta_B^{\text{obs}}$.


\subsubsection*{Gravitino or neutralino dark matter}

\begin{figure}
\centering
\includegraphics[width=0.7\textwidth]{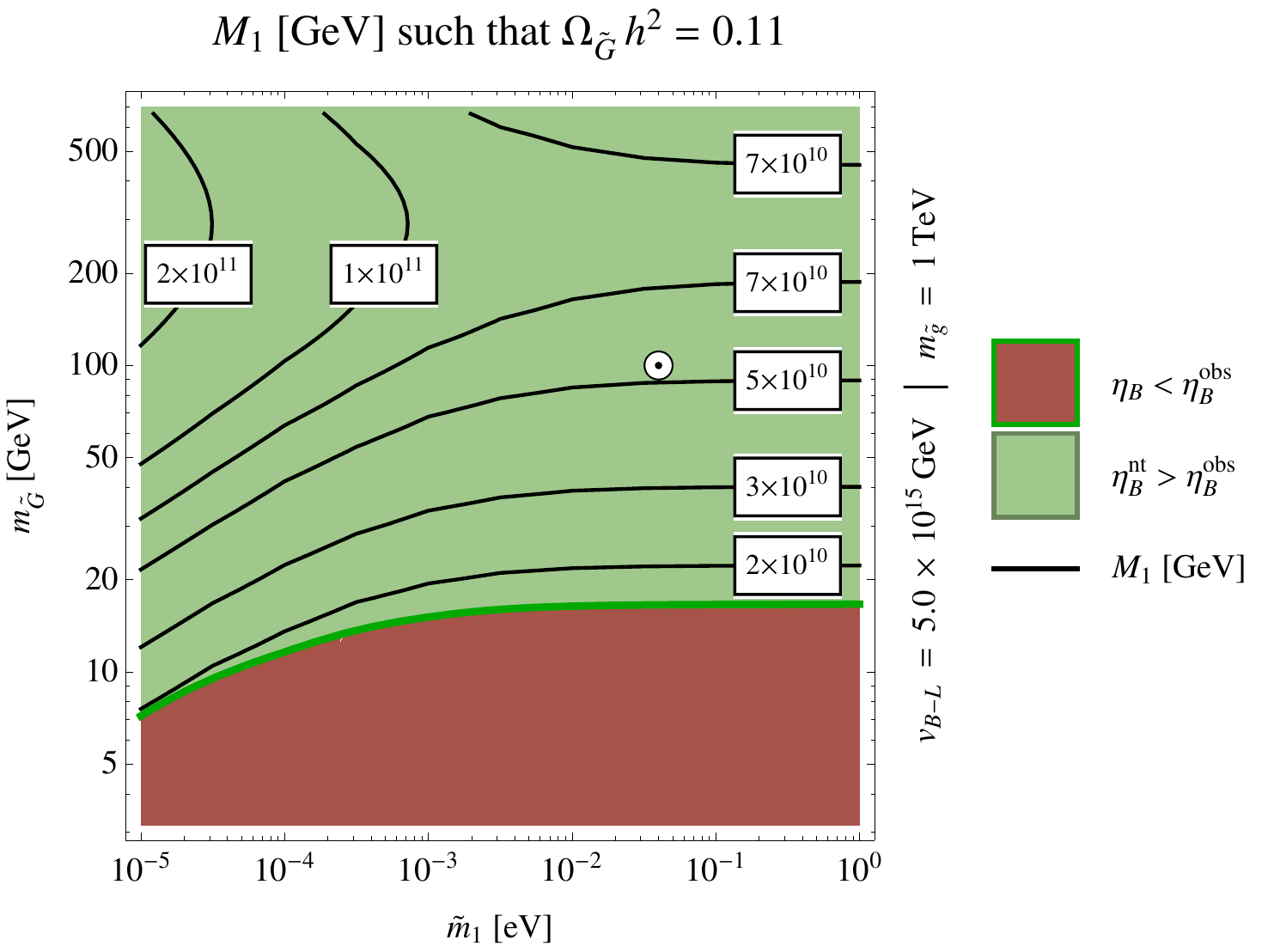}
\caption{Contour plots of the heavy neutrino mass $M_1$ as a function
of the effective neutrino mass~$\widetilde{m}_1$ and the gravitino mass
$m_{\widetilde{G}}$, such that the relic density of dark matter is
accounted for by gravitinos.
In the red region the lepton asymmetry generated by leptogenesis is
smaller than the observed one, providing a lower bound on the
gravitino mass depending on $\widetilde{m}_1$.
The small white circle marks the position of the parameter point
discussed in Figs.~\ref{fig:overview_numdens}--\ref{fig:BL_asymmetry}.
From Ref.~\cite{Buchmuller:2012wn}.
\label{fig:M1}}
\end{figure}

The thermal bath produced in the decays of the heavy neutrinos
gives rise to a thermal gravitino abundance, which can,
depending on the underlying low-energy sparticle spectrum,
be linked (either directly or via its decay products)
to today's dark matter abundance.
In the former case, we assume the gravitino to be the lightest
supersymmetric particle (LSP), as is, for instance, the case
in gauge or gaugino mediated scenarios of supersymmetry breaking.
We can then deduce today's gravitino dark matter
abundance, $\Omega_{\widetilde{G}} h^2$, from the final value
of the comoving gravitino abundance $N_{\widetilde G}$,
\begin{align}
\Omega_{\widetilde{G}} h^2 = \frac{\rho_{\widetilde{G}}^0}{\rho_c / h^2} =
\frac{m_{\widetilde{G}} \, n_\gamma^0}{\rho_c / h^{2}}
\frac{g_{*,s}^0}{g_{*,s}^\text{RH}} \left.\frac{N_{\widetilde{G}}}{N_\gamma}\right|_{a_f} \,,
\label{eq:OmegaGth2}
\end{align}
where $\rho_c = 3H^2/(8 \pi G) = 1.05 \times 10^{-5} \, h^2 \, \textrm{GeV}\, \textrm{cm}^{-3}$
denotes the critical energy density of the universe, $h$ the Hubble rate
in the units 
$H = h \times 100\,\textrm{km} \, \textrm{s}^{-1} \, \textrm{Mpc}^{-1}$ and
$n_\gamma^0 = 410\,\textrm{cm}^{-3}$ the present number density of the CMB photons.
Due to the high temperatures reached in our scenario, we do not expect
a significant contribution from nonthermal gravitino production.
For the parameter example shown in Fig.~\ref{fig:overview_numdens}, we find
$ \Omega_{\widetilde G} h^2 \simeq 0.11 \,,$
matching the observed amount of dark matter
$\Omega_{\text{DM}}^{\text{obs}} h^2 \simeq 0.11$~\cite{Komatsu:2010fb}%
\footnote{The recently published PLANCK data yields a slightly
larger value, $\Omega_\text{DM}^\text{obs} h^2 = 0.12$~\cite{Ade:2013zuv}.
The effect of this change on the work presented here is marginal, and in
the following we will stay with the value quoted above.}.
Note that, in the choice of this parameter example,
$M_1 = 5.4 \times 10^{11}$~GeV was adjusted to obtain this result.
Performing a parameter scan over $\widetilde m_1$ and
$m_{\widetilde G}$, while always adjusting $M_1$ so as to achieve
the correct gravitino dark matter abundance, yields the viable
parameter space of our model as depicted in Fig.~\ref{fig:M1}.
The red shaded region is excluded due to an insufficient
production of baryon asymmetry, whereas in the green shaded region
we produce a sufficient amount of baryon asymmetry (mainly nonthermally)
as well as the correct dark matter abundance.
In this region, the reheating temperature ranges
from ${\cal O}\left(10^8\right)$ to ${\cal O}\left(10^{10}\right)$~GeV.


As can be seen from Fig.~\ref{fig:M1}, requiring successful
leptogenesis as well as the correct dark matter abundance thus
yields a lower bound on the gravitino mass $m_{\widetilde G}$ in
terms of the effective neutrino mass parameter $\widetilde m_1$,
\begin{align}
m_{\widetilde{G}} \geq  16\,\textrm{GeV} \left(\frac{m_{\tilde g}}{1\,\textrm{TeV}}\right)^2
\left(\frac{\widetilde{m}_1}{10^{-3}\,\textrm{eV}}\right)^{0.25-c}
\,, \quad c =  \begin{cases}  -0.01 \quad \text{for  }
\widetilde m_1 \lesssim 10^{-3}~\text{eV} \\ \phantom{-}0.21 \quad \text{for  }
\widetilde m_1 \gtrsim 10^{-3}~\text{eV} \end{cases} \,.
\label{eq_mgravitino}
\end{align}
with the value of the exponent $c$ determined by numerically
solving the Boltzmann equations.
Eq.~\eqref{eq_mgravitino} links
a parameter of the neutrino mass sector related to $B$$-$$L$
breaking to a parameter involved in low-energy supersymmetry breaking.
Physically, this bound can be understood as follows.
For gravitino masses below $\mathcal{O}\left(10\right)\,\textrm{GeV}$,
a reheating temperature
$T_{\textrm{RH}}^N \lesssim \mathcal{O}\left(10^{8} \cdots 10^9\right)\,\textrm{GeV}$
is required to avoid overproduction of gravitinos.
According to our reheating mechanism, such low reheating
temperatures are associated with relatively small values of the
neutrino mass, $M_1 \lesssim \mathcal{O}\left(10^{10}\right)\,\textrm{GeV}$.
The low temperature and low mass then entail a small abundance of (s)neutrinos
at the time the asymmetry is generated and a small $CP$ parameter $\epsilon_1$.
Both effects combine and result in an insufficient lepton asymmetry,
rendering dark matter made of gravitinos with a mass below
$\mathcal{O}\left(10\right)\,\textrm{GeV}$ inconsistent with leptogenesis.


Alternatively, as discussed in Ref.~\cite{Buchmuller:2012bt}, we can assume a
hierarchical supersymmetric mass spectrum with the gravitino as the heaviest
particle and a neutralino with mass $m_\chi$ as the LSP,
\begin{align}
 m_\chi \ll m_\text{squark, slepton} \ll m_{\widetilde G}\,,
\label{eq:HiErArChY}
\end{align}
as is found, for instance, in Refs.~\cite{Ibe:2011aa,Jeong:2011sg,Krippendorf:2012ir}.
Due to this hierarchy, the LSP is typically a `pure' gaugino or
higgsino~\cite{Martin:1997ns}. Generically, the thermal abundance
of a bino LSP is too large.
We therefore focus on the possibility of a wino or higgsino LSP%
\footnote{Recently, it has been shown that wino DM is strongly constrained
by indirect searches using the H.E.S.S.\ and Fermi gamma-ray
telescopes~\cite{Cohen:2013ama,Fan:2013faa}.\label{fn:WiNoCoNsTrAiNtS}}.
There are then two relevant production channels for neutralino dark matter:
thermal production, accompanied by the standard thermal freeze-out mechanism
for weakly interacting massive particles (WIMPs), and nonthermal production,
as a decay product of the gravitinos produced during the reheating process.
In the parameter regime of interest, the resulting thermal and nonthermal
abundances can be estimated as
$\Omega_{\chi}^{\text{th}}$~\cite{ArkaniHamed:2006mb, Hisano:2006nn, Cirelli:2007xd}
and $\Omega_\chi^{\widetilde G}$, respectively:
\begin{align}
\begin{split}
\Omega^{\mathrm{th}}_{\chi} h^2 & = c_{\chi} 
\left(\frac{m_{\chi}}{1~\mathrm{TeV}}\right)^2 \ , \quad
\quad c_{\widetilde{w}} = 0.014\ , \quad c_{\widetilde{h}} = 0.10 \, \\
\Omega_\chi^{\widetilde G} h^2 & = \left( \frac{m_\chi}{m_{\widetilde G}} \right)
\Omega_{\widetilde G} h^2 \simeq 2.7\times 10^{-2} 
\left(\frac{m_{\chi}}{100~\mathrm{GeV}}\right)
\left(\frac{T_{\mathrm{RH}}^N (M_1,\widetilde{m}_1)}{10^{10}~\mathrm{GeV}} 
\right) \,.
\end{split}
\end{align}
Here, $c_{\widetilde w}$ and $c_{\tilde h}$ apply to the wino and higgsino
case, respectively, and $\Omega_{\widetilde G}$ refers to the `would-be'
gravitino abundance today if the gravitinos were stable.


\begin{figure} 
\centering
\includegraphics[width=0.7\textwidth]{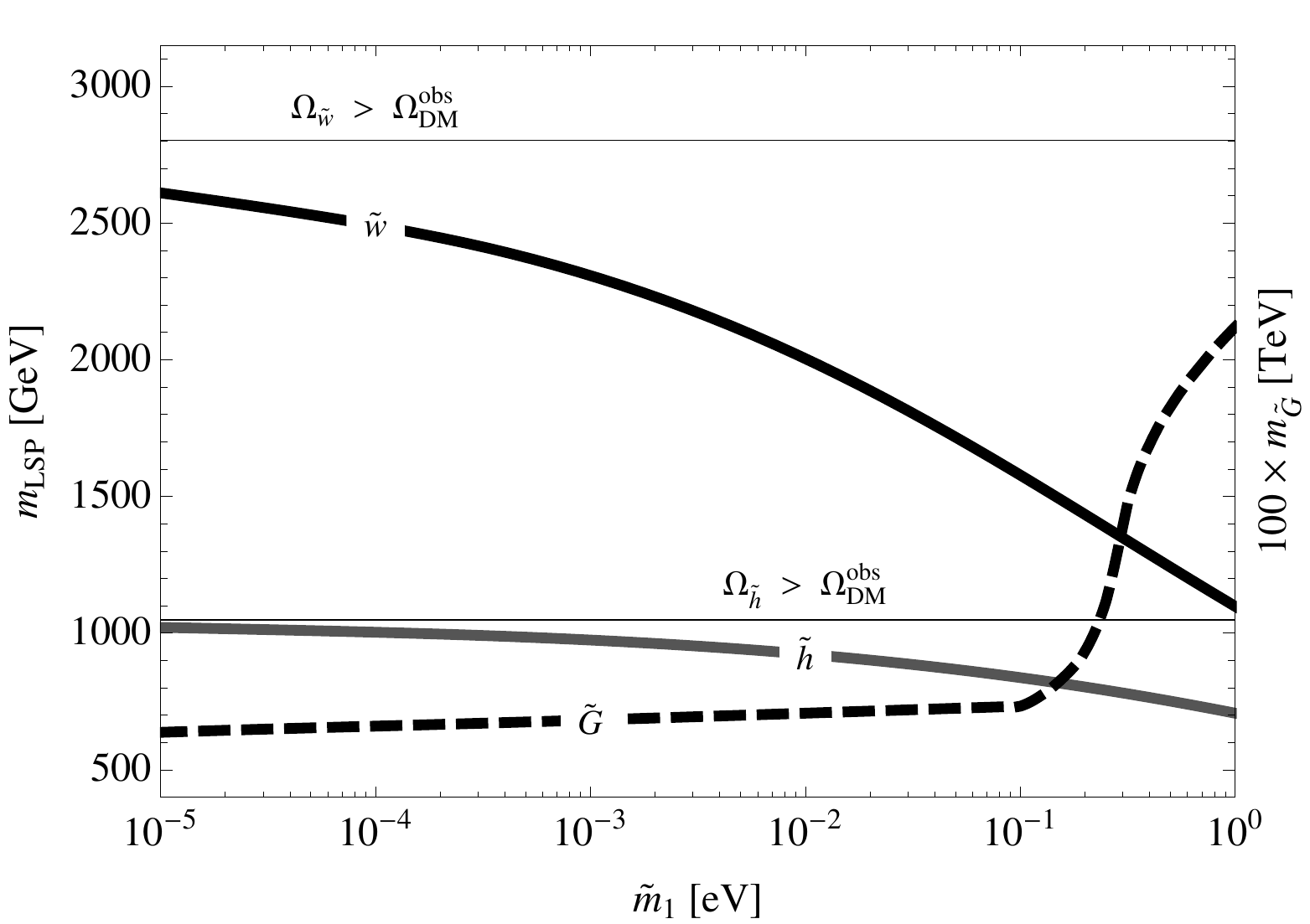}
\caption{Upper bounds on wino ($\widetilde{w}$) and higgsino
($\widetilde{h}$) LSP masses imposed by successful leptogenesis
as well as absolute lower bound on the gravitino mass according to BBN
as functions of the effective neutrino mass $\widetilde{m}_1$.
Wino masses larger than $2.8\,\textrm{TeV}$ and higgsino masses larger than
$1.0\;\textrm{TeV}$ result in thermal overproduction of DM.
From Ref.~\cite{Buchmuller:2012bt}.}
\label{fig_neutralino}
\end{figure}


Requiring the total neutralino LSP abundance to match the
observed dark matter abundance constrains the reheating temperature,
depending on the value of the neutralino LSP mass $m_\chi$.
Additionally taking into account the bounds from successful leptogenesis
and big bang nucleosynthesis (BBN) on the reheating temperature, we find
upper bounds on the neutralino LSP mass and an absolute lower bound on the
gravitino mass (for all neutralino LSP masses) depending on the value
of $\widetilde m_1$, as depicted in Fig.~\ref{fig_neutralino}.
As in the gravitino LSP case, we thus find relations between the
neutrino and superparticle mass spectrum, induced by the key role
of the reheating temperature in the efficiency of both leptogenesis
and thermal gravitino production.


\section{Gravitational Waves}
\label{sec_GW}

 
So far, we have discussed the birth of the hot early universe in the
MSM as well as indirect probes of this mechanism in terms of the
resulting neutrino and dark matter properties.
We now turn to the possibility of directly probing such early universe
physics by measuring the gravitational wave (GW) background~\cite{Allen:1996vm}.
GWs are generated by nonspherical, inhomogeneous strong gravitational field dynamics,
decouple immediately from their source and, to very good approximation,
propagate freely ever after.
Hence, GWs act as messengers carrying information on the early universe to us. 


\subsection{Cosmic gravitational wave background}
\label{sec:GWs}


Gravitational waves are tensor perturbations of the homogeneous background metric.
In a flat FRW background, these perturbations can be parametrized as~\cite{Maggiore:1900zz}
\begin{align}
 ds^2 = a^2(\tau) \, (\eta_{\mu \nu} + h_{\mu \nu}) dx^\mu dx^\nu \,.
\label{eq_metric}
\end{align}
Here, $\eta_{\mu \nu} = \text{diag}(-1, 1, 1, 1)$ accounts for the flat background
and $h_{\mu \nu}$ denotes the tensor perturbation around this background.
$x^\mu$ are conformal coordinates with $x^i$, $i = 1..3$, denoting the comoving
spatial coordinates and $\tau = x^0$ the conformal time.
These are related to the physical coordinates and the cosmic time
as $\boldsymbol x_\text{phys} = a(\tau)\, \boldsymbol x$ 
and $dt = a(\tau)\, d\tau$, respectively.


The dynamical evolution of the tensor perturbation
is described by the Einstein equation.
In the vacuum, $h_{\mu \nu}$ contains two physical degrees of freedom.
A convenient gauge choice is the transverse traceless (TT) gauge, 
i.e.\ $h^{0 \mu} = 0$, $h^i_i = 0$ and  $\partial^j h_{ij} = 0$. 
In the weak field approximation, the linearized Einstein equation in momentum
space  yields the following mode equation for the tensor perturbation around
the FRW background in the TT gauge,
\begin{align}
 \tilde{h}^{''}_{ij}(\boldsymbol{k}, \tau) + 
\left(k^2 - \frac{a^{''}}{a}\right) \tilde{h}_{ij}(\boldsymbol{k}, \tau) 
= 16 \pi \, G \, a \, \Pi_{ij}(\boldsymbol k, \tau) \,,
\label{eq_einstein_k}
\end{align}
describing the generation and propagation of GWs. Here $\tilde{h}_{ij} = a h_{ij}$,
$\Pi_{ij}$ denotes the Fourier transform of the TT part
of the anisotropic stress-energy tensor $T_{\mu  \nu}$ of the source, $k = |\boldsymbol k|$,
$\boldsymbol k$ is the comoving wave number, related to the physical wave number through 
$\boldsymbol k_\text{phys} = \boldsymbol k / a$, and the prime denotes
the derivative with respect to conformal time.


A useful plane wave expansion for freely propagating GWs is given by
\begin{align}
h_{ij}\left(\boldsymbol{x},\tau\right) = \sum_{P = +,\times}
\int_{-\infty}^{+\infty} \frac{dk}{2\pi} \int d^2\hat{\boldsymbol{k}} \, \,
h_P\left(\boldsymbol{k}\right) \, T_k(\tau) \,\,
e_{ij}^P\big(\hat{k}\big) \, 
e^{-ik\left(\tau - \hat{\boldsymbol{k}}\boldsymbol{x}\right)}\,,
\label{eq:hFourier}
\end{align}
where $\hat{\boldsymbol{k}} = \boldsymbol{k} / k$,
$P = +, \times$ labels the two possible polarization states of
a GW in the TT gauge and $e_{ij}^{+,\times}$ are the two corresponding
polarization tensors satisfying the normalization condition
$e_{ij}^P e^{ij\,Q} = 2 \delta^{PQ}$.
$h_P(\boldsymbol{k})$ denote the coefficients of the expansion after factorizing
out the redshift due to the expansion, with the latter captured
in the so-called transfer function $T_k(\tau)$.


An analytical expression for $T_k$ can be obtained by studying the
source-free version of Eq.~\eqref{eq_einstein_k}.
The resulting mode equation can be easily solved, revealing that 
the amplitude $h_{ij}(k)$ of a given mode remains constant in the
super-horizon regime, $k \ll aH$,  while it decreases as $1/a$ inside
the horizon, i.e.\ for $k \gg aH$. 
Identifying the transfer function $T_k$ as 
$T_k(\tau_*, \tau) =  h^E_{ij}(\boldsymbol{k}, \tau)/h^E_{ij}(\boldsymbol{k}, \tau_*)$,
with $h^E_{ij}(\boldsymbol{k}, \tau)$ denoting the envelope of the
oscillating function $h_{ij}(\boldsymbol{k}, \tau)$, we can employ the
approximation,%
\footnote{In Sec.~\ref{sec_reheating}, we set $a_\text{PH}=1$.
Another convention used frequently is $a_0 = 1$, with $a_0$ referring
to the value of the scale factor today.
In this section, we explicitly keep $a_0$ without specifying a
convention.
In the end, the dependence on $a_0$ must drop out of our expression for any
observable, irrespectively of the adopted convention.}
cf.\ e.g. Ref.~\cite{Smith:2005mm},
\begin{align}
T_k(\tau_*, \tau_0) \approx \frac{a(\tau_*)}{a(\tau_0)} \qquad \text{with  }
\tau_* = \begin{cases} \tau_i \text{  for sub-horizon sources} \\
\tau_{k} \text{  for super-horizon sources} \end{cases} \,.
\label{eq:approx_transferfunction}
\end{align}
Here, $\tau_i$ marks the time when the GW was generated and $\tau_{k}$ denotes
the time when a given mode with wave number~$k$ entered the horizon,
$k = a(\tau_{k}) \, H(\tau_{k}) \,$.
In Eq.~\eqref{eq:approx_transferfunction}, we assume for super-horizon
sources that the amplitude is constant until $\tau = \tau_{k}$ and then
drops as $1/a$ immediately afterwards.
The actual solution to the mode equation yields corrections
to both of these assumptions.
However, as a numerical check reveals, these two effects roughly
compensate each other, so that Eq.~\eqref{eq:approx_transferfunction}
reproduces the full result very well.
For super-horizon sources, we will use the more compact
notation $T_k(\tau) = T_k(\tau_k,\tau)$ in the following.


The GW background is a superposition of GWs propagating
with all frequencies in all directions.
An important observable characterizing the GW background
is the ensemble average of the energy density~\cite{Maggiore:1900zz}, which
is expected to be isotropic,
\begin{align}
\label{eq:rhoGWtot}
\rho_\text{GW}(\tau)  = \frac{1}{32 \pi G} 
\left \langle \dot h_{ij}\left(\boldsymbol{x},\tau\right) 
\dot h^{ij}\left(\boldsymbol{x},\tau\right) \right \rangle 
 = \int_{-\infty}^\infty d \ln k \,\, \frac{\partial \rho_\text{GW}(k, \tau)}
{\partial \ln k} \ ,
\end{align}
with the angular brackets denoting the ensemble average and the dot referring 
to the derivative with respect to cosmic time.
Alternatively, one uses the ratio of the differential energy density
to the critical density,
\begin{align}
\label{eq:OmegaGW}
\Omega_\text{GW}(k, \tau) = \frac{1}{\rho_c} \frac{\partial
\rho_\text{GW}(k, \tau)}{\partial \ln k} \ .
\end{align}
In the model considered in this paper, the energy density has partly
a quantum origin and partly a classical origin,
$\rho_\text{GW}(\tau) = \rho_\text{GW}^{\rm qu}(\tau) + \rho_\text{GW}^{\rm cl}(\tau)$.
The former part is due to inflation and is therefore stochastic, whereas the
latter part is determined by the contributions to the stress energy tensor
from cosmic strings and from tachyonic preheating,
$\rho^{\rm cl}_\text{GW}(\tau) = \rho_\text{GW}^{\rm CS} + \rho_\text{GW}^{\rm PH}(\tau)$.


For a stochastic GW background, the Fourier modes $h_A\left(\boldsymbol{k}\right)$
are random variables and their ensemble average of their
two-point function is determined by a time-independent spectral
density $S_h(k)$~\cite{Maggiore:1900zz},
\begin{align}
\left<h_P\left(\boldsymbol{k}\right)h_Q^*\left(\boldsymbol{k}'\right)\right>
= 2\pi \, \delta\left(k-k'\right) \frac{1}{4\pi} \, 
\delta^{(2)}\big(\hat{\boldsymbol{k}}-\hat{\boldsymbol{k}}'\big)
\, \delta_{PQ} \, \frac{1}{2} \, S_h(k) \ .
\label{eq:Shdef}
\end{align}
This relation reflects the fact that different modes are uncorrelated and
that the background is isotropic. 
Exploiting Eqs.~\eqref{eq:hFourier} through \eqref{eq:Shdef}, we can
express the differential energy density due to a stochastic
source in terms of the spectral density as
\begin{align}
\frac{\partial\rho_{\textrm{GW}}\left(k,\tau\right)}{\partial \ln k} =
\frac{a^2(\tau_*)}{16\pi^2 \, G \, a^4(\tau)} \, k^3 \, S_h(k) \,.
\label{eq:rhoGW}
\end{align}


The classical contribution to the GW energy density is obtained by integrating
Eq.~(\ref{eq_einstein_k}) from the initial time $\tau_i$ of GW production
until today,
\begin{align}
 h_{ij}(\boldsymbol{k}, \tau) = 
16 \pi G  \, \frac{1}{a(\tau)} \int_{\tau_i}^{\tau} d\tau' \; a(\tau') \, 
\mathcal{G}(k,\tau,\tau') \, \Pi_{ij}(\boldsymbol k, \tau') \,,
\label{eq_hcl_k}
\end{align}
where $\mathcal{G}(k,\tau,\tau')$ is the retarded Green's function of the
differential operator on the left-hand side of Eq.~(\ref{eq_einstein_k}). 
For sub-horizon modes, i.e.\ $k\tau \gg 1$, one has 
$\mathcal{G}(k,\tau,\tau') = \sin (k(\tau - \tau')) /k$.
With this, one can evaluate the ensemble average $\langle {\dot h}^2 \rangle$
in terms of $\langle \Pi^2 \rangle$ by calculating the derivative
of Eq.~\eqref{eq_hcl_k} on sub-horizon scales.
Assuming translation invariance and isotropy of the source,
\begin{align}
\left<\Pi_{ij}(\boldsymbol k, \tau) \Pi^{ij}(\boldsymbol k', \tau')\right> =
(2\pi)^3 \, \Pi^2(k, \tau, \tau') \, \delta(\boldsymbol k + \boldsymbol k')\,,
\label{eq_Pi_isotropy}
\end{align}
the resulting differential energy density simplifies to
\begin{align}\label{eq:GWsource}
\frac{\partial\rho_{\textrm{GW}}\left(k,\tau\right)}{\partial \ln k} =
\frac{2G}{\pi}\frac{ k^3}{a^4(\tau)} \int_{\tau_i}^{\tau} d\tau_1
\int_{\tau_i}^{\tau} d\tau_2 \;
a(\tau_1) \, a(\tau_2) \, \cos (k(\tau_1 - \tau_2)) \, \Pi^2(k,\tau_1,\tau_2)\,,
\end{align}
Here, in order to perform the ensemble average, we have also averaged the integrand 
over a period $\Delta\tau = 2\pi/k$, assuming ergodicity.


\subsection[Gravitational waves from a \texorpdfstring{$B$$-$$L$}{B-L} phase transition]
{Gravitational waves from a \texorpdfstring{\boldmath{$B$$-$$L$}}{B-L} phase transition}

\begin{figure}
\centering
\includegraphics[width=0.7\textwidth]{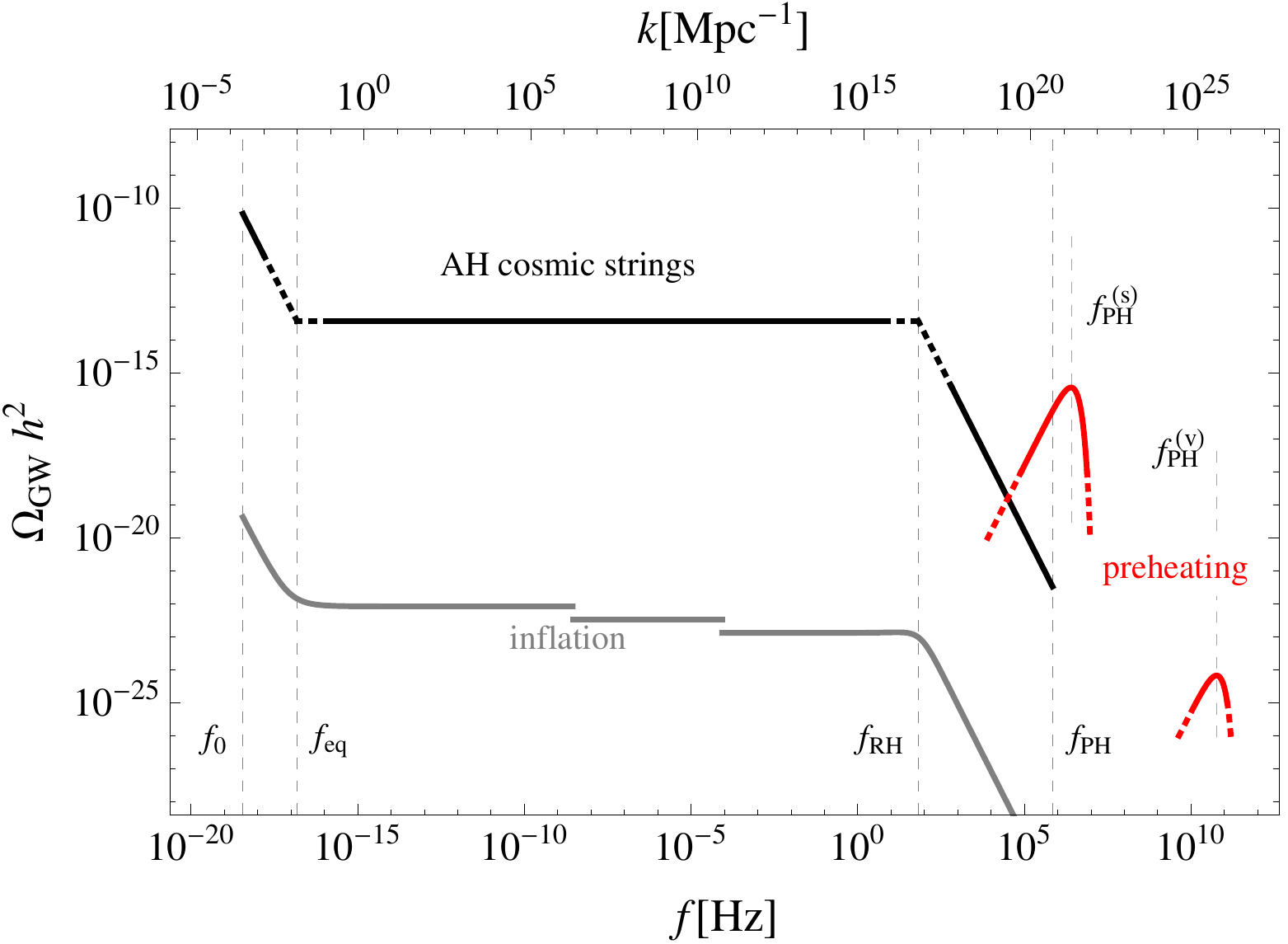}
\caption{Predicted GW spectrum due to inflation (gray), preheating (red)
and  Abelian Higgs cosmic strings (black) for $M_1~=~5.4~\times~10^{10}$~GeV,
$v_{B-L} = 5 \times 10^{15}$~GeV and $m_S = 3 \times 10^{13}$~GeV,
as in Fig.~\ref{fig:overview_numdens}.
$f_0$, $f_\text{eq}$, $f_\text{RH}$ and $f_\text{PH}$ denote the
frequencies associated with a horizon-sized wave today, at matter-radiation
equality, at reheating and at preheating, respectively.
$f_{\text{PH}}^{(s)}$ and $f_{\text{PH}}^{(v)}$ denote the positions
of the peaks in the GW spectrum associated with the scalar and
the vector boson present at preheating.
The dashed segments indicate the uncertainties due to the breakdown
of the analytical approximations.
From Ref.~\cite{Buchmuller:2013lra}.}
\label{fig:GW_overwiew}
\end{figure}

We will now discuss in turn the resulting GW background from inflation,
tachyonic preheating and cosmic strings in the scaling regime,
based on the analysis of Ref.~\cite{Buchmuller:2013lra}.
An overview of the resulting contributions is depicted in Fig.~\ref{fig:GW_overwiew}.


\subsubsection*{Gravitational waves from inflation}


During inflation, quantum fluctuations of the metric are generated and
stretched to ever larger physical scales so that they eventually cross
the Hubble horizon and become classical.
Outside the horizon, the amplitudes of these metric perturbations
remain preserved and they only begin to evolve again once they re-enter
the Hubble horizon after the end of inflation.
Inflation hence gives rise to a stochastic background of gravitational
waves~\cite{Turner:1993vb, Smith:2005mm, Nakayama:2008wy}  with a spectrum
which is determined by the properties of the primordial quantum metric
fluctuations as well as by the expansion history of the universe, which
governs the redshift of the GWs since horizon re-entry,
\begin{align}
 \Omega_{\textrm{GW}}(k,\tau) =  
\frac{A_t}{12} \, \frac{k^2}{a_0^2 H_0^2} \,T_k^2(\tau) \,.
\end{align}
Here, $A_t$, controlled by the Hubble parameter during inflation, denotes
the amplitude of the primordial tensor perturbations.
Evaluating the evolution of the scale factor throughout the cosmic
history, i.e.\ through the epochs of reheating, radiation, matter
and vacuum domination, yields the transfer function $T_k$ and thus
\begin{align}
\Omega_{\textrm{GW}} (k) = \frac{A_t^2}{12} \, \Omega_r \,
\frac{g_{*}^k}{g_{*}^0}
\left(\frac{g_{*,s}^0}{g_{*,s}^k}\right)^{4/3} \times
\begin{cases}
\frac{1}{2}\, \left(k_{\textrm{eq}} / k\right)^2 \,, & k_0 \ll k \ll k_{\textrm{eq}} \\
1 \,, & k_{\textrm{eq}} \ll k \ll k_{\textrm{RH}} \\
2\,R\, C_{\textrm{RH}}^6 \, \left(k_{\textrm{RH}} / k\right)^2 \,, &
k_{\textrm{RH}} \ll k \ll k_{\textrm{PH}}
\end{cases} \,,
\label{eq_inflation_result}
\end{align}
with $\Omega_r$ denoting the fraction of energy stored in radiation today.
The  parameters $C_\text{RH}$ and $R$ account for the deviation from pure
matter domination during reheating and the production of relativistic degrees
of freedom after $a_\text{RH}$, respectively, and are numerically found to
be typically ${\cal O}(1)$.%
\footnote{For a more detailed discussion of the numerical results, including
the precise shape of the `kinks' in the inflationary GW spectrum,
cf.\ Ref.~\cite{Buchmuller:2013lra}.}
As long as a mode with wave number $k$ re-enters the Hubble horizon
during radiation domination, $g_{*}^k$ and $g_{*,s}^k$ denote the usual
values of the effective number of degrees of freedom $g_{*}(\tau)$ and
$g_{*,s}(\tau)$ at time $\tau_k$.
On the other hand, during reheating and matter domination $g_{*}^k$
and $g_{*,s}^k$ correspond to
$g_{*}^{\textrm{RH}}$ and $g_{*,s}^{\textrm{RH}}$ as well as to
$g_{*}^{\textrm{eq}}$ and $g_{*,s}^{\textrm{eq}}$,
respectively.
The wave numbers $k_{\text{eq}}$, $k_\text{RH}$ and $k_{\text{PH}}$ refer
to the modes which crossed the horizon at matter-radiation equality, the end
of reheating and at preheating, respectively.
$k_0$ is correspondingly given by the size of the Hubble horizon today.
Translated into frequencies $f = k/(2 \pi a_0)$ at which GW experiments
could observe the corresponding modes, they are given by
\begin{align}
f_0 = & \:
3.58 \times 10^{-19} \,\textrm{Hz}&&\hspace{-2.2cm} \left(\frac{h}{0.70}\right) \,, \\
f_{\textrm{eq}} = & \:
1.57 \times 10^{-17} \,\textrm{Hz}&&\hspace{-2.2cm}
\left(\frac{\Omega_mh^2}{0.14}\right) \,, \label{eq:feq} \\
f_{\textrm{RH}} = & \:
4.25 \times 10^{-1} \,\textrm{Hz}&&\hspace{-2.2cm}
\left(\frac{T_*}{10^7\,\textrm{GeV}}\right) \,,  \label{eq:fRH} \\
f_{\textrm{PH}} = & \:
1.93 \times 10^4 \,\textrm{Hz}&&\hspace{-2.2cm}
\left(
\frac{\lambda}{10^{-4}}\right)^{1/6}
\left(\frac{10^{-15} \, v_{B-L}}{5 \,\textrm{GeV}}\right)^{2/3}
\left(\frac{T_*}{10^7\,\textrm{GeV}}\right)^{1/3} \,,
\end{align}
with $\Omega_m$ denoting the present value of the fraction of energy
stored in matter and $T_*$ closely related to the reheating temperature,
cf.\ footnote~\ref{footnote_temp}.
Evidently, the energy spectrum $\Omega_{\textrm{GW}}$ decreases
like $k^{-2}$ at its edges and features a plateau in its center,
cf.\ gray curve in Fig.~\ref{fig:GW_overwiew}.
In the context of cosmological $B$$-$$L$ breaking, the height of the
plateau is controlled by the coupling $\lambda$,
which determines the self-interaction 
of the $B$$-$$L$ breaking Higgs field, as well as by the $B$$-$$L$ scale $v_{B-L}$,
\begin{align}
\Omega^\text{pl}_{\textrm{GW}} h^2
&= 3.28 \times 10^{-22} \left(\frac{\lambda}{10^{-4}}\right)
\left(\frac{v_{B-L}}{5\times 10^{15}\ \textrm{GeV}}\right)^4
\left(\frac{\Omega_r}{8.5\times 10^{-5}}\right)\bar{g}^k \ ,
\end{align}
where $\bar{g}^k = (4 g_{*}^k/427)(427/(4 g_{*,s}^k))^{4/3}$ is a ratio
of energy and entropy degrees of freedom. 
The small steps visible in the plateau of the gray curve in
Fig.~\ref{fig:GW_overwiew} represent the change of the number of
relativistic degrees of freedom due to the QCD phase transition and  
the crossing of a typical mass-scale for supersymmetric particles.
A remarkable feature of the GW spectrum from inflation is that the
position of the kink, which separates the plateau arising for modes
which entered during radiation domination and the $k^{-2}$ behaviour
from the reheating regime, is directly related to the reheating temperature,
providing a possibility to probe this otherwise experimentally hardly
accessible quantity.%
\footnote{To be more precise, the quantity which is probed is
the temperature $\hat T_\text{RH}$ when the energy stored in
relativistic degrees of freedom (MSSM particles and nonthermal
(s)neutrinos) overcomes the energy stored in the nonrelativistic
$B$$-$$L$ Higgs bosons.
The quantity $T_*$ appearing in Eq.~\eqref{eq:fRH} is related to
$\hat T_{\rm RH}$ via two correction factors $D$ and $R$,
$T_* = R^{1/2} D^{1/3} \hat T_\text{RH}$.
Here $D$ accounts for the entropy production after $a = a_\text{RH}$ and,
just as $R$, it is typically found to be ${\cal O}(1)$ by numerically solving
the Boltzmann equations. \label{footnote_temp}}


\subsubsection*{Gravitational waves from preheating}


The process of tachyonic preheating acts as a classical, sub-horizon source
for GWs, which is active only for a short time.
The resulting GW spectrum can be obtained by calculating the solution
to the mode equation, Eq.~\eqref{eq_hcl_k}, and inserting it
into Eq.~\eqref{eq:rhoGWtot}.
The anisotropic stress tensor $\Pi_{ij}$ entering Eq.~\eqref{eq_hcl_k}
is determined by the dynamics of preheating and vanishes after the end
of preheating, allowing the GWs to propagate freely for
$\tau \gg \tau_{\text{PH}}$.
The remaining challenge is thus to calculate $\Pi_{ij}$ during
preheating.
This task can be performed numerically, cf.\ e.g.~Ref.~\cite{Dufaux:2007pt}
for a detailed description of the
method and an application to preheating after chaotic inflation
as well as Ref.~\cite{Dufaux:2008dn} for an application to
tachyonic preheating after hybrid inflation.
Based on analytical estimates supported by the results of these
simulations~\cite{Felder:2006cc,GarciaBellido:2007dg,Dufaux:2007pt,Dufaux:2008dn,Dufaux:2010cf},
one finds two high-frequency peaks in the resulting GW spectrum,
related to the mass of the $B$$-$$L$ vector ($v$) and
scalar Higgs ($s$) bosons at preheating.
The corresponding positions and amplitudes of the peaks
in the GW spectrum are given by
\begin{align}
 \begin{split}
f_{\text{PH}}^{(s)} & \simeq 6.3 \times 10^6 \text{ Hz }
\left( \frac{M_1}{10^{11} \text{ GeV}} \right)^{1/3} \left(\frac{5\times10^{15} \text{ GeV}}{v_{B-L}} \right)^{2}
\left( \frac{m_S}{3   \times    10^{13} \text{ GeV}} \right)^{7/6} , \\
\Omega_\text{GW}^{(s, \text{max})} h^2 & \simeq 3.6 \times 10^{-16}\  \frac{c_{\text{PH}}}{0.05} 
\left(\frac{M_1}{10^{11} \text{ GeV}} \right)^{  4/3}  \left(\frac{5\times10^{15} \text{ GeV}}{v_{B-L}}\right)^{-2}
\left(\frac{ m_S}{3   \times   10^{13} \text{ GeV}} \right)^{  -4/3} , \\
f_{\text{PH}}^{(v)} & \simeq 7.5 \times 10^{10} \text{ Hz }  g \left( \frac{M_1}{10^{11} \text{ GeV}} \right)^{1/3}
\left( \frac{m_S}{3   \times    10^{13} \text{ GeV}} \right)^{-1/2} , \\
\Omega_\text{GW}^{(v, \text{max})} h^2 & \simeq 2.6 \times 10^{-24} \, \frac{1}{g^2} \, \frac{c_{\text{PH}}}{0.05}
\left( \frac{M_1}{10^{11} \text{ GeV}} \right)^{  4/3} \left(\frac{5\times10^{15} \text{ GeV}}{v_{B-L}} \right)^{2}
\left( \frac{m_S}{3   \times   10^{13} \text{ GeV}} \right)^{2},
\end{split}
\label{eq_pred_preheating}
\end{align}
and are depicted by the red curves in Fig.~\ref{fig:GW_overwiew}.
Here $g$ is the $B$$-$$L$ gauge coupling and $c_\text{PH}$ is a
model-dependent numerical factor,
found to be $c_\text{PH} = 0.05$ in Ref.~\cite{GarciaBellido:2007dg}.


\subsubsection*{Gravitational waves from cosmic strings}


We now turn to the third source of GWs related to the $B$$-$$L$ phase transition:
cosmic strings in the scaling regime, cf.\ Sec.~\ref{sec_cosmic_strings}.
We here review the calculation of the resulting GW background in
the Abelian Higgs (AH) model following Ref.~\cite{Figueroa:2012kw}.
In Ref.~\cite{Buchmuller:2013lra}, we additionally discuss an
alternative approach based on the Nambu-Goto model of cosmic strings.
Here, we will merely give the final result of the latter calculation
in order to quantify the theoretical uncertainties involved. 


The GW background generated by an AH string network can be estimated
analytically starting from Eq.~\eqref{eq:GWsource}.
Exploiting general properties of the unequal time correlator of a scaling,
sub-horizon source as discussed in Ref.~\cite{Durrer:1998rw} and introducing
the dimensionless variable $x = k \tau$, we can evaluate the unequal
time correlator of the AH string network, $\Pi^2(k, \tau, \tau')$, as
\begin{align}
\Pi^2(k, \tau, \tau') = \frac{4 v_{B-L}^4}{\sqrt{\tau \tau'}} \, C^T(x, x')\ .
\label{eq_Pi_AH}
\end{align}
Here, $C^T(x, x')$ is essentially local in time \cite{Durrer:1998rw},
$C^T(x,x') \sim \delta(x-x') \, \widetilde C(x)$, with $\widetilde C$ some function
which rapidly falls off for $x \gg 1$, i.e.\ for modes well inside the horizon.
Inserting this into Eq.~\eqref{eq:GWsource} yields
\begin{align}
\Omega_\text{GW}(k) =  \frac{k^2}{3 \pi^2 H_0^2 a_0^2} 
\left(\frac{v_{B-L}}{M_{\rm Pl}}\right)^4 \int_{x_i}^{x_0} dx \;
\frac{a^2(x/k)}{a_0^2 \, x} \, \widetilde C(x) \,. 
\label{eq_OmegaGW_2}
\end{align}
As a result of the rapid decrease of $\widetilde C(x)$ for $x \gg 1$,
this integral is dominated by its lower boundary.
For scales which entered the Hubble horizon after the $B$$-$$L$ phase
transition, $x_i = k \, \tau_k$ is an ${\cal O}(1)$ constant.
Hence, the $k$-dependence of Eq.~\eqref{eq_OmegaGW_2} can be traced
back to $a(x/k)$.
For radiation domination, we have $a(\tau) \simeq \sqrt{\Omega_r} H_0 \tau a_0^2$,
where we have neglected the change in the effective number of degrees of freedom.
This yields
\begin{align}
\int_{x_i}^{x_0} dx \; \frac{a^2(x/k)}{a_0^2 \, x} \, \widetilde C(x) 
\simeq \frac{\Omega_r \, H_0^2 \, a_0^2}{2 \, k^2} \,  F^r \ ,
\end{align}
where $F^r$ is a constant, and therefore a flat spectrum, 
$\Omega_\text{GW} \propto k^0$.
For matter domination, one has $a(x/k) \propto k^{-2}$,
which yields $\Omega_\text{GW} \propto k^{-2}$. 
In summary, we can express today's spectrum of GWs from a
scaling network of AH cosmic strings as%
\footnote{Note that in Eq.~\eqref{eq_masterformula},
the normalization of the `$1/k^2$ flanks' was obtained by matching
to the plateau value for $k = k_{\text{RH}}$ and $k = k_{\text{eq}}$,
respectively.
However, since close to these points the dominant component of the
energy density is not much larger than the other components,
a more detailed knowledge of $\widetilde C(x)$ is necessary to
evaluate Eq.~\eqref{eq_OmegaGW_2} at these points. \label{footnote_normalization}}
\begin{align}
\Omega_\text{GW}(k) \simeq  \Omega^\text{pl}_\text{GW} \times
\left\{\begin{array}{cl} 
(k_{\rm eq}/k)^2, & \quad k_0 \ll k \ll k_{\rm eq} \\ 
1, & \quad k_{\rm eq} \ll k \ll k_{\rm RH}  \\
(k_{\rm RH}/k)^2 , & \quad  k_{\rm RH} \ll k \ll k_\text{PH}  \end{array}\right. \,.
\label{eq_masterformula}
\end{align}
Here, $k_{\rm eq}$, $k_{\rm RH}$ and $k_\text{PH}$ are determined
by Eqs.~\eqref{eq:feq}, and \eqref{eq:fRH},
and the height of the plateau $\Omega_\text{GW}^\text{pl}$
can be estimated using the result of the numerical
simulations in Ref.~\cite{Figueroa:2012kw},
\begin{align}
\Omega_\text{GW}^\text{pl} h^2
& = \: 4.0 \times 10^{-14} \, \frac{F^r}{F^r_{\text{FHU}}} 
\left( \frac{v_{B-L}}{5 \times 10^{15} \text{GeV}} \right)^4 
\left( \frac{\Omega_r h^2} {4.2 \times 10^{-5}} \right) \,,
\label{eq_omega_plat}
\end{align}
where $F^r_{\text{FHU}} = 4.0\times 10^3$ is the numerical constant
determined in Ref.~\cite{Figueroa:2012kw} for global cosmic strings.
The corresponding constant for local strings is
expected to have the same order of magnitude~\cite{hindmarsh}.


Eq.~\eqref{eq_masterformula} strikingly resembles the result found for
the stochastic GW background from inflation, cf.\ Eq.~\eqref{eq_inflation_result},
up to an overall normalization factor, cf.\ Fig.~\ref{fig:GW_overwiew}.
Note, however, that the origin is quite different.
On the one hand, in the case of inflation, the GWs can be traced back to
vacuum fluctuations of the metric, which remain `frozen' outside the horizon.
After horizon re-entry, they propagate according to the source-free wave
equation in FRW space.
The amplitude of the resulting stochastic GW background today
is determined by the redshift of these modes after entering the horizon.
On the other hand, the GWs from cosmic strings stem from a classical
source, which is active until today.
Only the nature of the unequal time correlator, with its rapid
decrease for $x \gg 1$, effectively removes the impact of the source when
the corresponding mode is well within the horizon.
In more physical terms, this implies that the dominant source for GWs 
from cosmic strings are Hubble-sized structures of the cosmic
string network.
This explains why the wave numbers associated with the horizon at
$a_{\text{RH}}$ and $a_{\text{eq}}$ play crucial roles in the GW
spectrum from AH cosmic strings, although the GW modes associated
with cosmic strings never actually `cross' the horizon.
For cosmic strings, the height of the plateau is enhanced by a
very large numerical factor~$F^r$.
On the contrary, GWs from inflation are suppressed by the
small Yukawa coupling~$\lambda$.
This explains the enormous difference in amplitude
between GWs from inflation and cosmic strings. 


\begin{figure}
\centering
\includegraphics[width=0.7\textwidth]{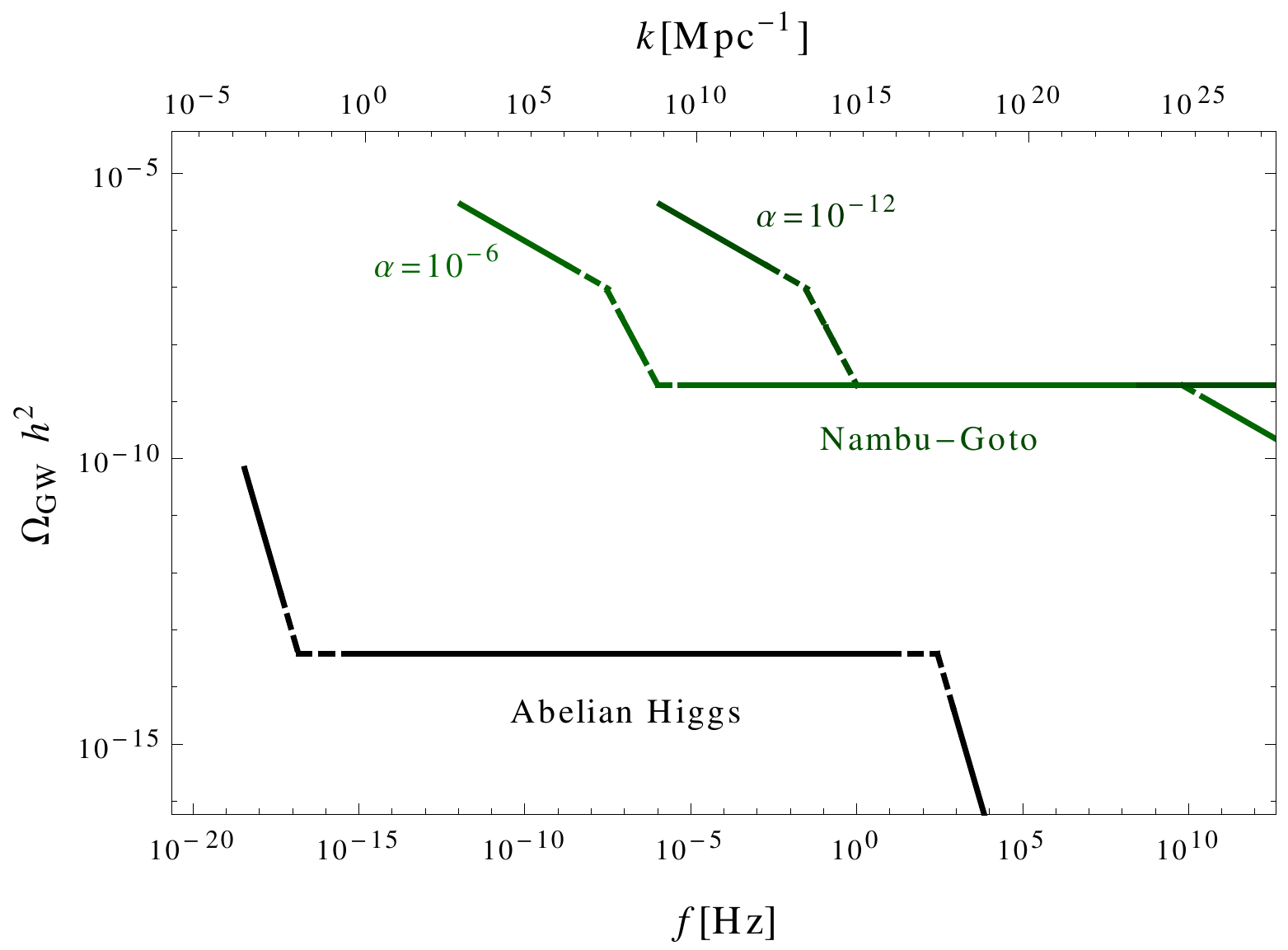}
\caption{Comparison of the GW spectra predicted by AH strings and
NG strings for two values of $\alpha$ (which governs the initial
cosmic string loop size in the NG model).
The other parameters are chosen as in Fig.~\ref{fig:GW_overwiew}, which
yields a cosmic string tension of $G\mu = 2\times 10^{-7}$.
From Ref.~\cite{Buchmuller:2013lra}.
\label{fig:comparison}}
\end{figure}


The calculation presented here, resulting in Eq.~\eqref{eq_masterformula},
was based on the Abelian Higgs (AH) cosmic string model.
For comparison, Fig.~\ref{fig:comparison} shows the result
obtained in the Nambu-Goto (NG) model, cf.\ Ref.~\cite{Buchmuller:2013lra}.
In both approaches the radiation-dominated epoch leads to a plateau for
intermediate frequencies.
Compared to the AH result, the boundaries in the NG case are
shifted to higher frequencies by a factor $1/\alpha$, where $\alpha$
denotes the size at which cosmic string loops are formed relative
to the respective horizon size.%
\footnote{Note that $\alpha$ cannot take arbitrarily small values.
A lower bound is given by the requirement that, in the validity range
of the NG model, the loop size should be larger
than the string width obtained in the AH model (controlled by
$m_S^{-1}, m_G^{-1}$) or at the very least larger than $M_{\rm Pl}^{-1}$.} 
This shift in frequency is directly related to the maximal loop
size which is determined by $\alpha H^{-1}$ in the NG case.
Furthermore, the frequency dependence for small and large
frequencies differs, which is a consequence of the different
mechanisms of gravitational radiation:
in the AH model the dominant contribution to the GW background
comes from Hubble-sized structures, in the NG model the dominant
contribution is due to `cusps' in small cosmic string loops,
which are formed when waves moving in opposite directions on
the loop collide.
The striking difference in amplitude by five orders of magnitude
between the AH and NG model is due to the different energy loss mechanisms
of the string network in the scaling regime.
While the energy loss of AH strings is mainly due to massive radiation,
NG strings deposit all their energy into GWs.
Hence, these two cases provide lower and upper bounds on the GW
background from cosmic strings, and it is conceivable that the true answer
corresponds to some intermediate value.
Assuming a transition between the AH model at early times and
the NG model at later times sometime during radiation domination,
a notable point is that,  due to the shift of the GW spectrum of
NG strings to higher frequencies, the GWs generated at later times
in the NG regime might cover up the GWs generated at earlier times in
the AH regime.
Properly addressing this important question of how to correctly describe
the evolution of cosmic strings is clearly a theoretical challenge.


\section{Observational Prospects and Outlook}

 
\begin{figure}
\centering
\includegraphics[width=1\textwidth]{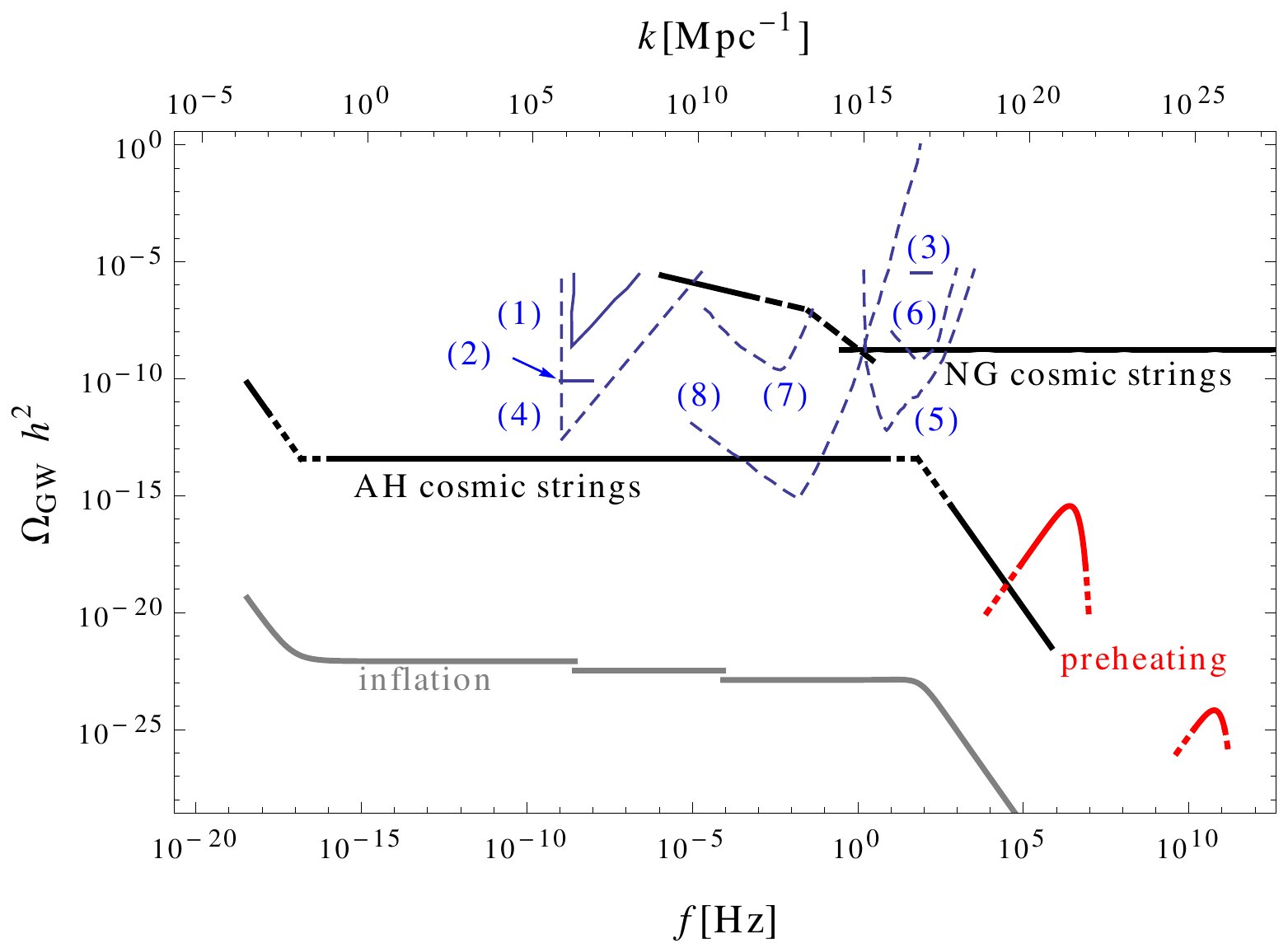}
\caption{Predicted GW spectrum and the (expected)
sensitivity of current and upcoming experiments.
The GW spectrum due to inflation (gray), preheating (red) as well as AH and NG cosmic
strings (black) is shown for $v_{B-L} = 5 \times 10^{15}\,\textrm{GeV}$,
$M_1 = 10^{11}\,\textrm{GeV}$, $m_S = 3\times10^{13}\,\textrm{GeV}$, and
$\alpha = 10^{-12}$.
The current bounds on the stochastic GW spectrum from (1) millisecond pulsar
timing (taken from Ref.~\cite{Smith:2005mm}), with (2) marking the update from
EPTA~\cite{vanHaasteren:2011ni} and (3) LIGO~\cite{Palomba:2012wn} are
marked by solid blue lines.
The dashed blue lines mark the expected sensitivity
of some planned experiments: (4) SKA~\cite{Kramer:2004rwa},
(5) ET~\cite{etreport}, (6) advanced LIGO~\cite{etreport},
(7) eLISA~\cite{AmaroSeoane:2012km}, (8) BBO and DECIGO~\cite{Alabidi:2012ex}.
Note that with a correlation analysis ultimate DECIGO has a
sensitivity down to $10^{-18}$. From Ref.~\cite{Buchmuller:2013lra}}
\label{fig:GWprospects}
\end{figure}


In this paper, we have demonstrated that the MSM, a minimal supersymmetric
extension of the Standard Model with right-handed neutrinos and spontaneously
broken local $B$$-$$L$ symmetry, is capable of remedying several
shortcomings of the Standard Model, while, at the same time, successfully
accounting for the earliest phases of the cosmological evolution.
While the MSM allows for grand unification and explains the
smallness of the observed neutrino masses on the particle physics side,
it accommodates inflation and the generation of entropy, baryon
asymmetry and dark matter during the reheating process on the cosmology side.
These successes of the MSM therefore truly render it a minimal
consistent model of particle physics \textit{and} the early universe.


The MSM gives rise to a rich phenomenology that can be probed in on-going
and upcoming cosmological observations and high energy physics experiments.
First of all, future data on the temperature anisotropies as well as
on the polarization of the CMB radiation will test the dynamics of the
$B$$-$$L$ breaking sector of the MSM.
Dedicated experiments searching for tensor modes in the CMB,
such as CMBPol~\cite{Baumann:2008aq} or LiteBIRD~\cite{LiteBIRD},
have, for instance, the potential to rule out supersymmetric F-term
hybrid inflation by measuring a tensor-to-scalar ratio $r$ of
$\mathcal{O}(10^{-2})$ or larger, cf.\ Eq.~\eqref{eq_predictions_inflation}.
Meanwhile, indications in the CMB for the presence of local cosmic strings
could provide evidence in favor of cosmological $B$$-$$L$ breaking,
cf.\ Sec.~\ref{sec_cosmic_strings}.
For parameter values compatible with inflation, the AH model of the
$B$$-$$L$ phase transition typically predicts a cosmic string tension
$G\mu$ only shortly below the current observational bound, $G\mu<3.2\times 10^{-7}$.
From the perspective of the MSM, it is thus expected that signs of
cosmic strings should be seen soon.


Next to the CMB, cosmic strings ought to reveal their existence also in weak and strong lensing
surveys, in the spectrum of ultra-high-energy cosmic rays and GeV-scale $\gamma$-rays
and finally also in the spectrum of GWs.
In Sec.~\ref{sec_GW}, we discussed this latter characteristic signature of the MSM in more detail.
In doing so, we put particular emphasis on the uncertainties in the theoretical calculations,
which we assessed by calculating the spectrum of GWs either emitted by AH or by NG cosmic strings.
Our result for the GW spectrum related to cosmological $B$$-$$L$ breaking is shown in
Fig.~\ref{fig:GWprospects}, which compares the GW signals that are respectively expected
to originate from AH strings, NG strings, inflation, and preheating.
In addition to that, Fig.~\ref{fig:GWprospects} also indicates current bounds on
$\Omega_{\textrm{GW}} h^2$ as well as the expected sensitivity of upcoming GW
experiments, cf.\ Ref.~\cite{Maggiore:1999vm} for a review.
The observation of a GW signal coming from cosmic strings
in the scaling regime in the not-too-far future is clearly challenging.
Depending on the parameters of the AH model, the reheating temperature and the
cosmic string loop parameter $\alpha$, future experiments will either see
a flat plateau in the GW spectrum or detect a kink-type feature related to the transition between
two successive stages in the expansion history of the universe.
Particularly intriguing in this context is the possibility to determine the reheating
temperature by measuring the position of the kink in the GW spectrum
that is induced by AH strings at frequencies around $f_{\textrm{RH}}$,
cf.\ Eq.~\eqref{eq:fRH} and footnote~\ref{footnote_temp}.
Nonetheless, it is important to realize that, at present, our
understanding of the formation, evolution
and decay of cosmic strings is still far from complete.
For one reason or another, the GW background due to cosmic strings might be suppressed or
even absent, cf.\ Ref.~\cite{Buchmuller:2013lra} for details,
thereby potentially rendering inflation and preheating
the dominant sources of GWs.
At least in the case of inflation, the exact shape of the GW spectrum
and in particular of its kinks could then be predicted with a much
better precision than as for cosmic strings~\cite{Buchmuller:2013lra}. 
As both the GW signals from inflation as well as from preheating are, however,
rather faint, a positive observation by any of the planned GW
experiments seems to be out of reach.
%

The dynamics of the neutrino sector in the MSM can be tested on the basis of the
parameter relations that we derived in our study of the reheating process,
cf.\ Sec.~\ref{subsec_reheating_pheno}.
Assuming the gravitino to be the LSP, the requirement of consistency between
leptogenesis and gravitino dark matter provided us with relations between
the neutrino mass parameters $\widetilde{m}_1$ and $M_1$ on the one hand
and the superparticle masses $m_{\widetilde{G}}$ and $m_{\tilde{g}}$ on the
other hand, cf.\ Fig.~8.
In particular, we found a lower bound on the gravitino mass that scales
quadratically with the gluino mass and that at the same time slightly
varies with $\widetilde{m}_1$, cf.\ Eq.~\eqref{eq_mgravitino}.
As an alternative to gravitino dark matter, we also considered the possibility
of very heavy gravitinos, in the case of which dark matter is accounted for
by partly thermally, partly nonthermally produced winos or higgsinos.
In this scenario, requiring consistency between leptogenesis, WIMP
dark matter and primordial nucleosynthesis, we were able to derive
upper bounds on the neutralino mass $m_{\textrm{LSP}}$ as well as
absolute lower bounds on the gravitino mass as functions of
$\widetilde{m}_1$, cf.\ Fig.~\ref{fig_neutralino}.


Owing to these relations and bounds, a determination of $\widetilde{m}_1$, $M_1$,
$m_{\widetilde{G}}$, $m_{\tilde{g}}$ and/or $m_{\textrm{LSP}}$ in present or upcoming
experiments would therefore allow to constrain the parameter space of the MSM or even to falsify it.
The absolute mass scale of the Standard Model neutrinos, which is closely related to
$\widetilde{m}_1$, is, for instance, probed by low-energy
neutrino experiments such as GERDA~\cite{Agostini:2013mzu} and KATRIN~\cite{fortheKATRIN:2013saa}
that are looking for neutrinoless double-$\beta$ decay and
studying the $\beta$-decay of tritium, respectively.
Meanwhile, it is hard to experimentally access the neutrino mass $M_1$ directly;
but fortunately the MSM offers a possibility to determine $M_1$ indirectly.
As reheating after inflation is driven by the decay of heavy (s)neutrinos in the MSM,
the plateau temperature $T_{\textrm{RH}}^N$ turns out to be a function of $\widetilde{m}_1$
and $M_1$, cf.\ Ref.~\cite{Buchmuller:2012wn} for details.
Once $\widetilde{m}_1$ is known, there thus exists a one-to-one relation between values
of $M_1$ and $T_{\textrm{RH}}^N$, at least as long as all Yukawa couplings are kept fixed.
As mentioned above, it is conceivable that the reheating temperature could possibly
be determined by means of GW observations.
Such an observation would then also allow for a measurement of $M_1$.


Depending on the scale of soft supersymmetry breaking and the details of the
superparticle mass spectrum, a determination of $m_{\widetilde{G}}$,
$m_{\tilde{g}}$ and $m_{\textrm{LSP}}$ is potentially within the reach of
experiments aiming at the direct or indirect detection of dark matter
and/or collider searches for supersymmetry.
If dark matter should be composed of gravitino LSPs, direct and indirect detection experiments
would actually be bound to yield null observations.
However, if $R$ parity was slightly violated, gravitino dark matter would be
unstable~\cite{Takayama:2000uz,Buchmuller:2007ui}, which could lead to observable signals in
the fluxes of gamma rays, charged cosmic rays and cosmic neutrinos~\cite{Grefe:2011dp}.
At the same time, the decays of the next-to-lightest superparticle (NLSP) via its
$R$ parity-violating interactions might be observable in collider experiments
in the form of displaced vertices with distinctive decay
signatures~\cite{Bobrovskyi:2011vx,Bobrovskyi:2012dc}.
A slight violation of $R$ parity is in fact motivated from cosmology---if $R$
parity was exactly conserved, the late-time decays of the NLSP could spoil
the successful predictions of primordial
nucleosynthesis~\cite{Moroi:1993mb,deGouvea:1997tn}---and hence we are confident
that the nature of dark matter is not doomed to remain obscure, even if it is made
out of gravitinos.
Finally, dark matter composed of partly thermally, partly nonthermally
produced winos or higgsinos could soon be seen in indirect detection
experiments such as H.E.S.S. and Fermi-LAT, cf.\ footnote~\ref{fn:WiNoCoNsTrAiNtS}.
On the other hand, for the hierarchical superparticle mass spectrum in
Eq.~\eqref{eq:HiErArChY}, the prospects
for a direct detection of WIMP dark matter via its
scattering off heavy nuclei do not look particularly
promising, cf.\ Ref.~\cite{Buchmuller:2012bt}.
Also the discovery of a wino or higgsino LSP at colliders seems to be rather
challenging in this scenario.
Given the large masses for the gluinos and squarks, the characteristic missing energy
signature of events with LSPs in the final state may be absent.
On the contrary, the wino- or higgsino-like chargino, almost mass degenerate with
its neutral partner, might leave macroscopic charged
tracks in the detector, which could increase the discovery potential of this dark matter scenario.
In addition to that, monojets caused by the Drell-Yan production of LSP pairs
in association with initial state gluon radiation may also provide a possible discovery channel.
We therefore conclude that the MSM is experimentally accessible not only
in cosmological observations, but also in a number of high energy
physics experiments.
Upcoming data will thus shed more light on whether or not the MSM
is indeed a good description of particle physics up to the 
multi-TeV scale as well as of the earliest phases of our universe.


Before concluding, we would still like
to compare the MSM with a closely related model, the $\nu$MSM~\cite{Shaposhnikov:2010zza},
the non-supersymmetric minimal Standard Model with
right-handed neutrinos, which can also account for inflation, entropy
production, baryon asymmetry and dark matter. 
The $\nu$MSM is a model with minimal particle content as well as minimal symmetry.
The local symmetry is that of the Standard Model and
the global $B$$-$$L$ symmetry is explicitly broken by Majorana masses
of the right-handed neutrinos.
Baryogenesis via leptogenesis and dark
matter require these masses to lie in the keV and GeV range, far below
the electroweak scale, which leads to predictions that are
experimentally testable in the near future.
The Higgs field of electroweak symmetry breaking also plays the
role of the inflaton, which requires a large non-minimal coupling to gravity, tuned to
account for the observed amplitude of the CMB
temperature anisotropies.
In the $\nu$MSM, there is no unification of strong and electroweak interactions.
Also the MSM has minimal matter content.
However, the symmetry group is enlarged, and in addition to the
Standard Model gauge group it contains local $U(1)_{B-L}$ symmetry and local
supersymmetry.
Assuming quark and lepton mass matrices compatible with grand unification,
and therefore hierarchical right-handed neutrinos,
one finds that $U(1)_{B-L}$ is broken at the GUT scale.
The symmetry breaking sector contains an inflaton field and the GUT scale
automatically yields the right order of magnitude for the amplitude of
CMB temperature anisotropies.
The lightest superparticle is the constituent of dark matter, which
can be searched for at the Large Hadron Collider as well as with
direct and indirect detection experiments.
Direct evidence for the MSM may eventually be obtained via the
spectrum of relic GWs.
In summary, supporting evidence for or falsification of the $\nu$MSM or
the MSM will decide whether or not physics beyond the Standard Model
is tied to symmetries larger than those already revealed by the Standard Model.


\subsubsection*{Acknowledgements}

This work has been supported in part by the German Science Foundation (DFG)
within the Collaborative Research Center 676 ``Particles, Strings and the Early Universe''
(W.\,B., V.\,D., K.\,K.) and by the World Premier International Research Center Initiative
(WPI) of the Ministry of Education, Culture, Sports, Science and Technology (MEXT) of Japan (K.\,S.).



\end{document}